\RequirePackage[mathlines,pagewise]{lineno}
\documentclass[prd,twocolumn,amsmath,amssymb,showpacs,superscriptaddress,nofootinbib]{revtex4-1}

\usepackage[english]{babel}
\usepackage{graphicx,epsfig}
\usepackage{dcolumn}
\usepackage{bm,color}
\usepackage{multirow}
\usepackage[dvipdfm,CJKbookmarks=true,unicode,colorlinks,linkcolor=blue,anchorcolor=blue,citecolor=blue,pdfborder={0 0 0}]{hyperref}

\graphicspath{{figures/}}

\usepackage{overpic,graphicx}
\usepackage{dcolumn}
\usepackage{bm}
\usepackage{rotating}
\usepackage{amsfonts}
\usepackage{keyval,graphicx}
\usepackage{textcomp,wasysym}

\usepackage{float}
\graphicspath{{figures/}}

\begin{document}

\hyphenation{BESII}
\hyphenation{BESIII}
\hyphenation{BEPCII}

\newcommand{\chisq}[1]{$\chi^{2}_{#1}$}
\newcommand{\GeV}{GeV/$c^2$}
\newcommand{\MeV}{MeV/$c^2$}
\newcommand{\br}[1]{\mathcal{B}(#1)}
\newcommand{\etap}{\eta^\prime}
\newcommand{\psip}{\psi^{\prime}}

\title{\Large \boldmath \bf Observation of Electromagnetic Dalitz decays $J/\psi \to P e^+e^-$}
\author{
\small
 M.~Ablikim$^{1}$, M.~N.~Achasov$^{8,a}$, X.~C.~Ai$^{1}$, O.~Albayrak$^{4}$, M.~Albrecht$^{3}$,
 D.~J.~Ambrose$^{41}$, F.~F.~An$^{1}$, Q.~An$^{42}$, J.~Z.~Bai$^{1}$, R.~Baldini Ferroli$^{19A}$
% ...
\\
\vspace{0.2cm}
(BESIII Collaboration)\\
\vspace{0.2cm} {\it
$^{1}$ Institute of High Energy Physics, Beijing 100049, People's Republic of China\\
% ...
$^{3}$ Bochum Ruhr-University, D-44780 Bochum, Germany\\
$^{4}$ Carnegie Mellon University, Pittsburgh, Pennsylvania 15213, USA\\
% ...
$^{8}$ G.I. Budker Institute of Nuclear Physics SB RAS (BINP), Novosibirsk 630090, Russia\\
% ...
$^{19}$ (A)INFN Laboratori Nazionali di Frascati, I-00044, Frascati, Italy; (B)INFN and University of Perugia, I-06100, Perugia, Italy\\
% ...
$^{41}$ University of Rochester, Rochester, New York 14627, USA\\
$^{42}$ University of Science and Technology of China, Hefei 230026, People's Republic of China\\
% ...
$^{a}$ Also at the Novosibirsk State University, Novosibirsk, 630090, Russia\\
% ...
}
\vspace{0.5cm}
}

\begin{abstract}
Based on a sample of $(225.3\pm2.8)\times 10^{6}$ $J/\psi$ events
collected with the BESIII detector, the electromagnetic Dalitz
decays of $J/\psi \to P e^+e^-(P=\etap/\eta/\pi^0)$ are studied. By
reconstructing the pseudoscalar mesons in various decay modes, the
decays $J/\psi \to \etap e^+e^-$, $J/\psi \to \eta e^+e^-$ and
$J/\psi \to \pi^0 e^+e^-$ are observed for the first time. The
branching fractions are determined to be $\mathcal{B}(J/\psi\to
\etap e^+e^-) = (5.81\pm0.16\pm0.31)\times10^{-5}$,
$\mathcal{B}(J/\psi\to \eta e^+e^-) =
(1.16\pm0.07\pm0.06)\times10^{-5}$, and $\mathcal{B}(J/\psi\to \pi^0
e^+e^-)=(7.56\pm1.32\pm0.50)\times10^{-7}$, where the first errors
are statistical and the second ones systematic.

\end{abstract}


\maketitle

\section{Introduction}\label{Introduction}
The study of electromagnetic (EM) decays of hadronic states plays an
important role in revealing the structure of hadrons and the
mechanism of the interactions between photons and
hadrons~\cite{Landsberg}. Notably, the EM Dalitz decays
$V\rightarrow P e^+e^-$ of unflavored vector ($V$) mesons ($V=\rho$,
$\omega$, $\phi$ or $J/\psi$) are of interest for probing the EM
structure arising at the vertex of the transition from vector to
pseudoscalar ($P$) states. In these decays, the lepton pair can be
formed by internal conversion of an intermediate virtual photon with
invariant-mass $M_{e^+e^-}$. Assuming point-like particles, the
variation of the decay rate with $M_{e^+e^-}$ is exactly described
by quantum electrodynamics (QED)~\cite{QED}. For physical mesons,
however, the rate will be modified by the dynamic transition form
factor  $|F_{VP} (q^2)|^2$, where $q$ is the total four-momentum of
the lepton pair and $q^2 = M^2_{e^+e^-}$ is their invariant-mass
squared. The general form for the $q^2$-dependent differential decay
width for $V \rightarrow P e^+e^-$, normalized to the width of the
corresponding radiative decay $V \rightarrow P \gamma$, is given
by~\cite{Landsberg}
\begin{widetext}
\begin{eqnarray}
 \frac{d\Gamma(V \rightarrow P e^+e^-)}{d q^2 \Gamma (V \rightarrow P \gamma)}
 &=&
 \frac{\alpha_{em}}{3\pi} |F_{V P}(q^2)|^2 \frac{1}{q^2} \left(1-\frac{4m^2_e}{q^2}\right)^{1/2}
\left(1+\frac{2m^2_e}{q^2}\right)
 \left[\left(1+\frac{q^2}{m^2_{V}-m^2_P}\right)^2 - \frac{4m^2_{V}q^2}{(m^2_{V}-m^2_P)^2}\right]^{3/2}\nonumber \\
  &=&  |F_{V P}(q^2)|^2 \times [\mbox{QED}(q^2)],
 \label{eq:dgamman}
\end{eqnarray}
\end{widetext}
where $m_V$ is the mass of the initial vector state, $m_P$ and $m_e$
are the masses of the final states pseudoscalar meson and lepton,
respectively; $\alpha_{em}$ is the fine structure constant, and
$[\mbox{QED}(q^2)]$ represents the point-like QED result. The
magnitude of the form factor can be estimated based on
phenomenological models of nonperturbative quantum chromodynamics
(QCD)~\cite{qcd1,qcd2,qcd3,qcd4,qcd5}. For example, in the vector
meson dominance (VMD) model~\cite{budnev}, the form factor is
governed mainly by the resonance interaction between photons and
hadrons in the time-like region. Experimentally, the form factor is
directly accessible by comparing the measured invariant-mass
spectrum of the lepton pairs from Dalitz decays with the point-like
QED prediction~\cite{QED}. In the simple pole
approximation~\cite{Becirevic,Becher} the $q^2$-dependent form
factor is parameterized by
\begin{eqnarray}\label{eq:simplepole}
   |F_{VP}(q^2)|=\frac{1}{(1-q^2/\Lambda^2)},
\end{eqnarray}
where the parameter $\Lambda$ is the spectroscopic pole mass.

The EM Dalitz decays of the light unflavored mesons $\rho, \omega$
and $\phi$ have been intensively studied by the CMD2, SND, NA60 and
KLOE experiments~\cite{UperRho, phimu,phiee,NA60,kloe}. For the
decays of $\phi\to\eta e^+e^-$ and $\omega \to\pi^0 e^+e^-$, the
branching fractions and slopes of the form factors $\Lambda^{-2}$
are measured~\cite{phimu,phiee,NA60,kloe} and the results are in
agreement with VMD predictions. Recently, however, a measurement of
$\omega\to\pi^0\mu^+\mu^-$ from the NA60 experiment~\cite{NA60}
obtains a value of $\Lambda^{-2}$ which is  ten standard deviations
from the expectations of VMD.

These theoretical and experimental investigations of the EM Dalitz
decays of the light vector mesons motivate us to study the rare
charmonium decays $J/\psi \rightarrow P e^+e^-$, which should provide
useful information on the interaction of the charmonium states with the
electromagnetic field. At present, there is no experimental
information on these decays. In Ref.~\cite{Jinlin}, by assuming a
simple pole approximation, the decay rates are estimated to be
$10^{-5}$ and $10^{-7}$ for the $J/\psi \rightarrow \etap (\eta)
e^+e^-$ and $\pi^0 e^+e^-$, respectively.  In this paper, we present
measurements of the branching fractions of $J/\psi \rightarrow P
e^+e^-$. This analysis is based on $(225.3\pm2.8)\times 10^{6}$
$J/\psi$ events~\cite{Jpsi}, accumulated with the Beijing
Spectrometer III (BESIII) detector~\cite{BESIII}, at the Beijing
Electron Positron Collider II (BEPCII).

\section{The BESIII experiment and Monte Carlo simulation} \label{Detector}

The BESIII detector and BEPCII accelerator represent major upgrades
over the previous versions, BESII and BEPC; the facility is used for
studies of hadron spectroscopy and $\tau$-charm physics. The design
peak luminosity of the double-ring $e^+e^-$ collider, BEPCII, is
$10^{33}$~cm$^{-2}$~s$^{-1}$ at a beam current of 0.93~A. The BESIII
detector has a geometrical acceptance of 93\% of 4$\pi$ solid angle
and consists of four main components; the inner three are enclosed
in a superconducting solenoidal magnet of 1.0~T magnetic field.
First, a small-celled, helium-based main drift chamber (MDC) with 43 layers
provides charged particle tracking and measurements of ionization energy
loss ($dE/dx$). The average single wire resolution is 135~$\mu$m, and the momentum
resolution for 1~GeV/$c$ charged particles is 0.5\%. Next is a
time-of-flight system (TOF) for particle identification (PID)
composed of a barrel part made of two layers with 88 pieces of 5~cm
thick, 2.4~m long plastic scintillators in each layer, and two end
caps with 96 fan-shaped, 5~cm thick, plastic scintillators in each
end cap. The time resolution is 80~ps in the barrel, and 110~ps in
the end caps, corresponding to a 2$\sigma$ K/$\pi$ separation for
momenta up to about 1.0 GeV/$c$. Third is an electromagnetic
calorimeter (EMC) made of 6240 CsI (Tl) crystals arranged in a
cylindrical shape (barrel) plus two end caps. For 1.0~GeV photons,
the energy resolution is 2.5\% in the barrel and 5\% in the end
caps, and the position resolution is 6~mm in the barrel and 9~mm in
the end caps. Finally, a muon chamber system made of 1272~m$^2$ of
resistive plate chambers arranged in 9 layers in the barrel and 8
layers in the end caps is incorporated in the return iron of the
superconducting magnet. The position resolution is about 2~cm.

Optimization of event selection and estimations of physical
backgrounds are performed using Monte Carlo (MC) simulated samples.
The \textsc{geant4}-based~\cite{GEANT4} simulation software BOOST
includes the geometric and material descriptions of the BESIII
detector, the detector response and digitization models, and also
tracks the detector running conditions and performance. The
production of the $J/\psi$ resonance is simulated by the MC event
generator \textsc{kkmc}~\cite{KKMC}; the known decay modes are
generated by \textsc{evtgen}~\cite{GEN, bes3gen} with branching
ratios set at the world average values~\cite{PDG}, while unknown
decays are generated by \textsc{lundcharm}~\cite{LUND}.
The analysis is performed in the framework of the BESIII offline
software system which takes care of the detector calibration,
event reconstruction and data persistency.

In this analysis, $J/\psi\to\etap e^+e^-$ is studied using
$\etap\to\gamma\pi^+\pi^-$ and $\etap\to\pi^+\pi^-\eta$ with
$\eta\to\gamma\gamma$; $J/\psi\to \eta e^+e^-$ is studied using
$\eta\to\gamma\gamma$ and $\eta\to\pi^+\pi^-\pi^0$ with
$\pi^0\to\gamma\gamma$; $J/\psi\to \pi^0 e^+e^-$ is studied using
$\pi^0\to\gamma\gamma$. An independent data sample of approximately
2.9 fb$^{-1}$ taken at $\sqrt{s}$=3.773 GeV is utilized to study
potential continuum background.

The \textsc{evtgen} package is used to generate
$J/\psi\to \etap e^+e^- $, $\eta e^+e^-$ and $\pi^0 e^+e^-$ events,
with angular distributions simulated according to the
amplitude squared in Eq.(3) of Ref.~\cite{Jinlin}.
A simple pole approximation is assumed for the form factor.
The decay $\eta\to\pi^+\pi^-\pi^0$ is generated according to the
Dalitz plot distribution measured in Ref.~\cite{EtaDalitz}. For the decay
$\etap \to\gamma \pi^+ \pi^-$, the generator takes $\rho$-$\omega$ interference
and box anomaly into account~\cite{GammaRhoDIY}, while the decay $\etap
\rightarrow \pi^+\pi^-\eta$ is generated with phase space.

\section{data analysis} \label{Analysis}

Charged tracks in the BESIII detectors are reconstructed from
ionization signals in the MDC. To select well-measured tracks we
require the polar angle to satisfy $|\cos\theta| < 0.93$ and that
tracks to pass within 10~cm of the interaction point in the beam
direction and within 1~cm in the plane perpendicular to the beam.
The number of such tracks and their net charge must exactly correspond
to the particular final state under study. For particle identification,
information from $dE/dx$ and TOF is combined to calculate the
probabilities, $\rm{Prob}_{\rm{PID}}($i$)$, that these measurements
are consistent with the hypothesis that a track is an electron,
pion, or kaon; $i = e, \pi, K$ labels the particle type. For both
electron and positron candidates, we require
$\rm{Prob}_{\rm{PID}}(e)>\rm{Prob}_{\rm{PID}}(\pi)$ and
$\rm{Prob}_{\rm{PID}}(e)>\rm{Prob}_{\rm{PID}}(K)$. The remaining
tracks are assumed to be pions, without PID requirements.

Electromagnetic showers are reconstructed from clusters of energy
depositions in the EMC crystals. The energy deposited in nearby TOF
counters is included to improve the reconstruction efficiency and
energy resolution. The shower energies are required to be greater
than 25~MeV for the barrel region $(|\rm{cos}(\theta) | < 0.80)$ and
50~MeV for the end cap region $(0.86 < |\rm{cos}(\theta)| < 0.92)$.
The showers in the angular range between the barrel and end cap are
poorly reconstructed and excluded from the analysis. To exclude
showers from charged particles, a photon candidate must be separated by at
least $10^{\circ}$ from any charged track.  Cluster timing
requirements are used to suppress electronic noise and energy
depositions unrelated to the event.

Events with the decay modes shown in Table~\ref{yields} are
selected. Every particle in the final state must be explicitly
found. For each mode, a vertex fit is performed on the charged
tracks; a loose $\chi^2$ cut ensures that they are consistent with
originating from the interaction point.
In $\etap/\eta$ channels with $\etap\rightarrow
\pi^+\pi^- \eta$ and $\eta \rightarrow \pi^+\pi^-\pi^0$, photon
pairs are used to reconstruct $\eta$ or $\pi^0$ candidates if the
invariant-mass satisfies $m_{\gamma\gamma} \in $ (480, 600)~\MeV~ or
(100, 160)~\MeV, respectively.
To improve resolution and reduce backgrounds, a four-constraint (4C)
energy-momentum conserving kinematic fit is performed.
For states with extra photon candidates, the combination with the least
$\chi^2_{\rm{4C}}$ is selected, and in all cases $\chi^2_{\rm{4C}}$
is required to be less than 100.

\begin{table}[hbtp]
  \centering
   \caption{For each decay mode, the number of observed signal events
   ($N_S$), the number of expected total peaking background events ($N_B$) in the signal
   region, and the MC efficiency ($\epsilon$) for signal are given. The uncertainty on $N_S$ is statistical only,
   and the signal regions are defined to be within $3\sigma$  of the nominal pseudoscalar masses.}
   \label{yields}\small
     \renewcommand{\arraystretch}{1.5}
     \begin{tabular}{lcccc}
        \hline\hline
           \hspace{0.2cm}   Modes   \hspace{0.2cm}        & \hspace{0.2cm} $N_{S}$ \hspace{0.2cm}  & \hspace{0.2cm}  $N_{B}$  \hspace{0.2cm} & \hspace{0.2cm} $\epsilon$\hspace{0.2cm} \\ \hline

        $J/\psi\to \etap e^+e^-(\etap\to \gamma\pi^+\pi^-)$      & $983.3\pm33.0$ & $27.4\pm1.0$  & 24.8\% \\

        $J/\psi\to \etap e^+e^-(\etap\to \pi^+\pi^-\eta)$        & $373.0\pm19.9$ & $8.5\pm0.3$ & 17.6\%  \\ \hline

        $J/\psi\to \eta e^+e^-(\eta\to \pi^+\pi^-\pi^0)$        & $84.2\pm9.6$   & $5.3\pm0.3$ & 14.9\%  \\

        $J/\psi\to \eta e^+e^- (\eta\to \gamma\gamma)$           & $235.5\pm16.4$ & $8.7\pm0.3$ & 22.7\%  \\ \hline

        $J/\psi\to \pi^0 e^+e^- (\pi^0\to \gamma\gamma) $         & $39.4\pm6.9$ & $1.1\pm0.1$ & 23.4\%    \\
        \hline\hline
      \end{tabular}
\end{table}

\begin{table}[hbtp]
  \centering
   \caption{ The normalized number of peaking background events ($N_{\gamma-\rm{conv}}$) from $J/\psi \rightarrow P \gamma$ with the
    photon subsequently converted into an electron-positron pair, and the corresponding MC efficiency ($\epsilon_{\gamma-\rm{conv}}$) for each background mode.}
   \label{gconv}\small
     \renewcommand{\arraystretch}{1.5}
     \begin{tabular}{lccc}
        \hline\hline
           \hspace{0.2cm}   Mode   \hspace{0.2cm}    & \hspace{0.2cm}  $N_{\gamma-\rm{conv}}$  \hspace{0.2cm} & \hspace{0.2cm} $\epsilon_{\gamma-\rm{conv}}$ \hspace{0.2cm} \\ \hline

        $J/\psi\to \etap \gamma(\etap\to \gamma\pi^+\pi^-)$     & $25.0\pm0.9$  & $7.4\times10^{-5}$ \\

        $J/\psi\to \etap \gamma(\etap\to \pi^+\pi^-\eta)$       & $7.6\pm0.3$ & $3.9\times10^{-5}$  \\ \hline

        $J/\psi\to \eta \gamma(\eta\to \pi^+\pi^-\pi^0)$       & $2.1\pm0.1$ & $3.7\times10^{-5}$   \\

        $J/\psi\to \eta \gamma(\eta \to \gamma\gamma)$          & $8.4\pm0.3$ & $8.6\times10^{-5}$   \\ \hline

        $J/\psi\to \pi^0 \gamma(\pi^0\to \gamma\gamma) $        & $0.7\pm0.1$ & $8.8\times10^{-5}$    \\
        \hline\hline
      \end{tabular}
\end{table}
\begin{figure}[htbp]
    \centering
        \includegraphics[width=4.3cm]{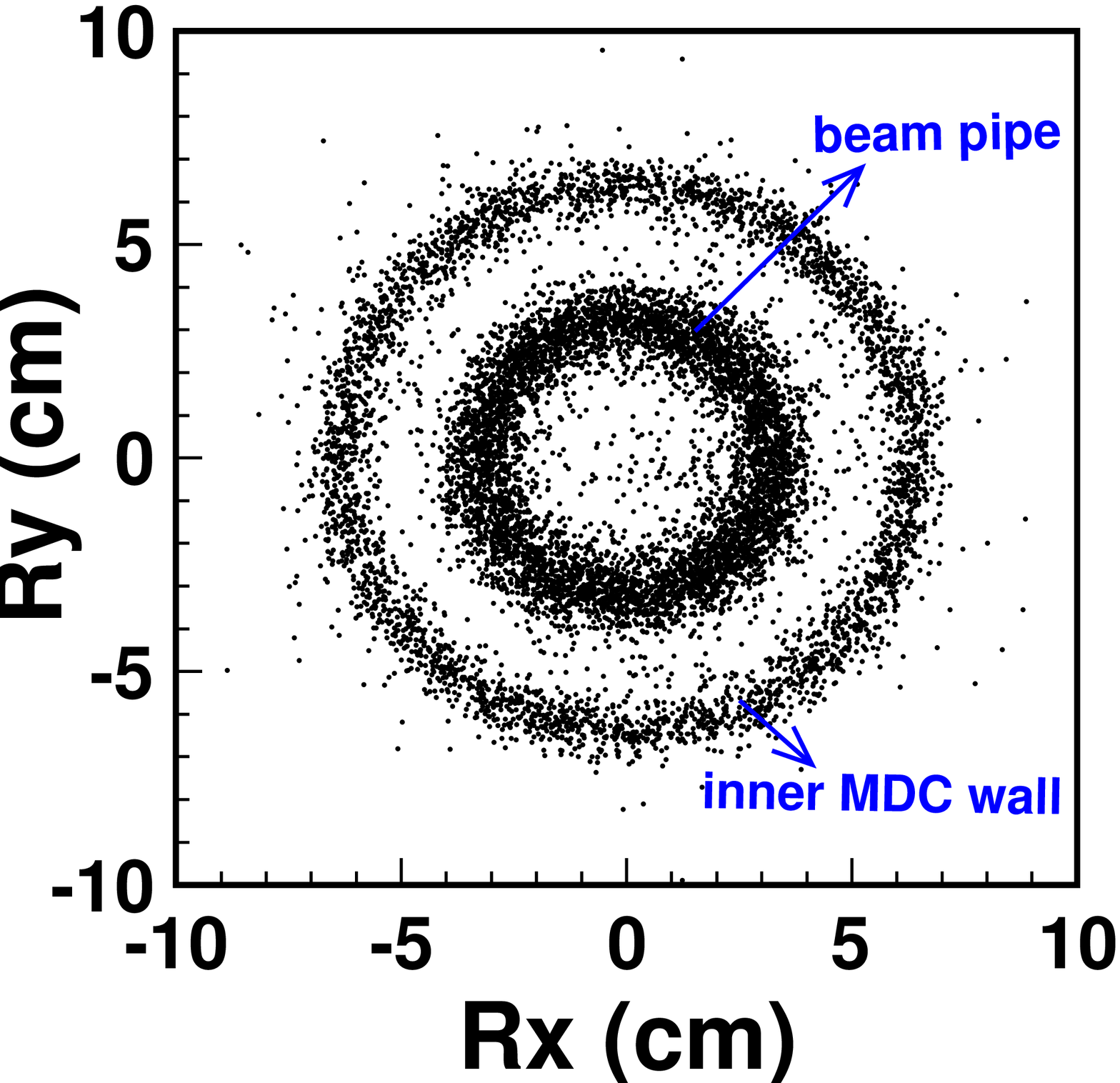}\put(-98,90){\bf \large~(a)}
        \includegraphics[width=4.3cm]{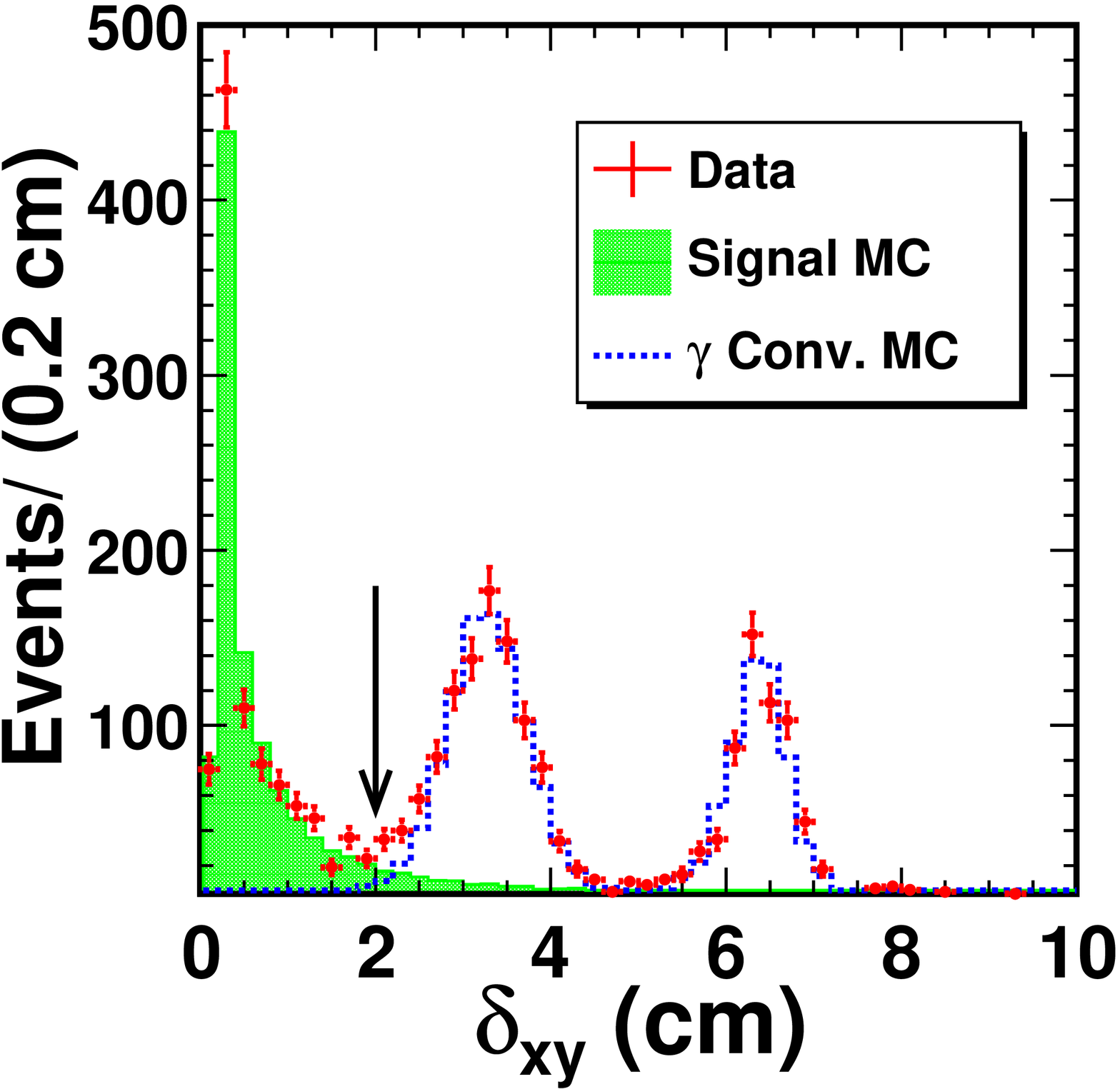}\put(-98,90){\bf \large~(b)}
        \caption{
        Veto of $\gamma$-conversion events. (a) a scatter plot of
        $R_{y}$ versus $R_{x}$ for the MC-simulated
        $J/\psi \rightarrow \etap \gamma$ ($\etap \rightarrow \gamma
        \pi^+\pi^-$) events. (b) $\delta_{xy}$ distributions. The (green)
shaded histogram shows the MC-simulated $J/\psi \rightarrow
e^+e^-\etap $ ($\etap \rightarrow \gamma \pi^+\pi^-$) signal events.
The (red) dots with error bars are data. The (blue) dotted histogram shows the
background from the $\gamma$-conversion events. In (b), the solid
arrow indicates the requirement on $\delta_{xy}$.}
    \label{rxy}
\end{figure}

In the analysis, one of the most important backgrounds comes from
events of the radiative decay $J/\psi \rightarrow P \gamma$ followed
by a $\gamma$ conversion in the material in front of the MDC,
including the beam pipe and the inner wall of the MDC.  To suppress
these backgrounds, a photon-conversion finder~\cite{GammaConv} was
developed to reconstruct the photon-conversion point in the
material.  The distance from this reconstructed conversion point to
the origin in the x-y plane, defined as
$\delta_{xy}=\sqrt{R^2_x+R^2_y}$, is used to distinguish photon
conversion background from signal; $R_x$ and $R_y$ are the distances
projected in the $x$ and $y$ directions, respectively. A scatter
plot of $R_y$ versus $R_x$ is shown in Fig.~\ref{rxy}(a) for the MC
simulated decay $J/\psi \rightarrow \etap \gamma$($\etap \rightarrow
\gamma \pi^+\pi^-$), in which one of the photons undergoes
conversion to an $e^+e^-$ pair. As indicated in Fig.~\ref{rxy}(a),
the inner circle matches the position of the beam pipe while the
outer circle corresponds to the position of the inner wall of the
MDC. Figure~\ref{rxy}(b) shows the $\delta_{xy}$ distributions for
the MC simulated $J/\psi \rightarrow \etap e^+e^-$ and $\etap
\gamma$ events, as well as the selected events  in the data for
comparison. In the $\delta_{xy}$ distributions, the two peaks above
2.0~cm correspond to the photon-conversion of the $\gamma$ from
$J/\psi \rightarrow \etap \gamma$ events in the material of the beam
pipe and inner wall of the MDC, while the events near
$\delta_{xy}=0$~cm are from the EM Dalitz decay. The selected events
from data are in good agreement with the MC simulations as shown in
Fig.~\ref{rxy}(b). Thus we require $\delta_{xy}<2$~cm to suppress
the photon-conversion backgrounds for all signal modes. This
requirement retains about 80\% of the signal events and removes
about 98\% of the photon-conversion events from the decay $J/\psi \rightarrow
\etap \gamma$. The ability of this requirement to veto the
photon-conversion events is the same for the other decay modes. The
normalized number of the peaking background events  from $J/\psi
\rightarrow P \gamma$ and the corresponding selection efficiencies
are listed in Table~\ref{gconv}.

In addition to  $J/\psi \to P \gamma$, further peaking backgrounds
arise from $J/\psi \rightarrow \phi P$, $\omega P$ and $\rho P$
($P=\etap$, $\eta$ or $\pi^0$) where $\phi$, $\omega$ and $\rho$
decay into $e^+e^-$. Studies based on MC simulations predict $2.2\pm
0.4$, $0.8\pm 0.1$, $2.8\pm 0.3$ and $0.4\pm0.1$ background events
for
$J/\psi\to \etap e^+e^-(\etap\to \gamma\pi^+\pi^-)$,
$J/\psi\to \etap e^+e^-(\etap\to \pi^+\pi^-\eta)$,
$J/\psi\to \eta e^+e^-(\eta\to \pi^+\pi^-\pi^0)$
and $J/\psi\to \pi^0 e^+e^-(\pi^0\to \gamma\gamma) $ modes, respectively.

Peaking background may also come from $J/\psi \rightarrow \pi^+\pi^- P$
with two pions misidentified as an $e^+ e^-$ pair.
The predicted background levels are 0.2, 0.1, 0.4, and 0.3 events
(with negligible errors) for
$J/\psi\to \etap e^+e^-(\etap\to \gamma\pi^+\pi^-)$,
$J/\psi\to \etap e^+e^-(\etap\to \pi^+\pi^-\eta)$,
$J/\psi\to \eta e^+e^-(\eta\to \pi^+\pi^-\pi^0)$,
and $J/\psi\to \eta e^+e^-(\eta\to \gamma\gamma)$, respectively.
For $J/\psi\to \pi^0 e^+e^-(\pi^0\to \gamma\gamma)$, the potential
peaking background from $J/\psi \rightarrow \pi^+\pi^-\pi^0$
(which has a large branching fraction of $(2.07\pm0.12)\%$~\cite{PDG})
is rejected by requiring $M_{e^+e^-}\le 0.4$ \GeV.
About 80\% of signal events are retained and the remaining background
is negligible.
Background from $J/\psi \rightarrow \phi P$
($\phi \rightarrow K^+K^-$) with two kaons misidentified as an
$e^+ e^-$ pair is also negligible based on the MC simulation.
The total expected peaking backgrounds from all sources
are summarized in Table~\ref{yields}.

\begin{figure}[htbp]
    \centering
        \includegraphics[width=4.4cm]{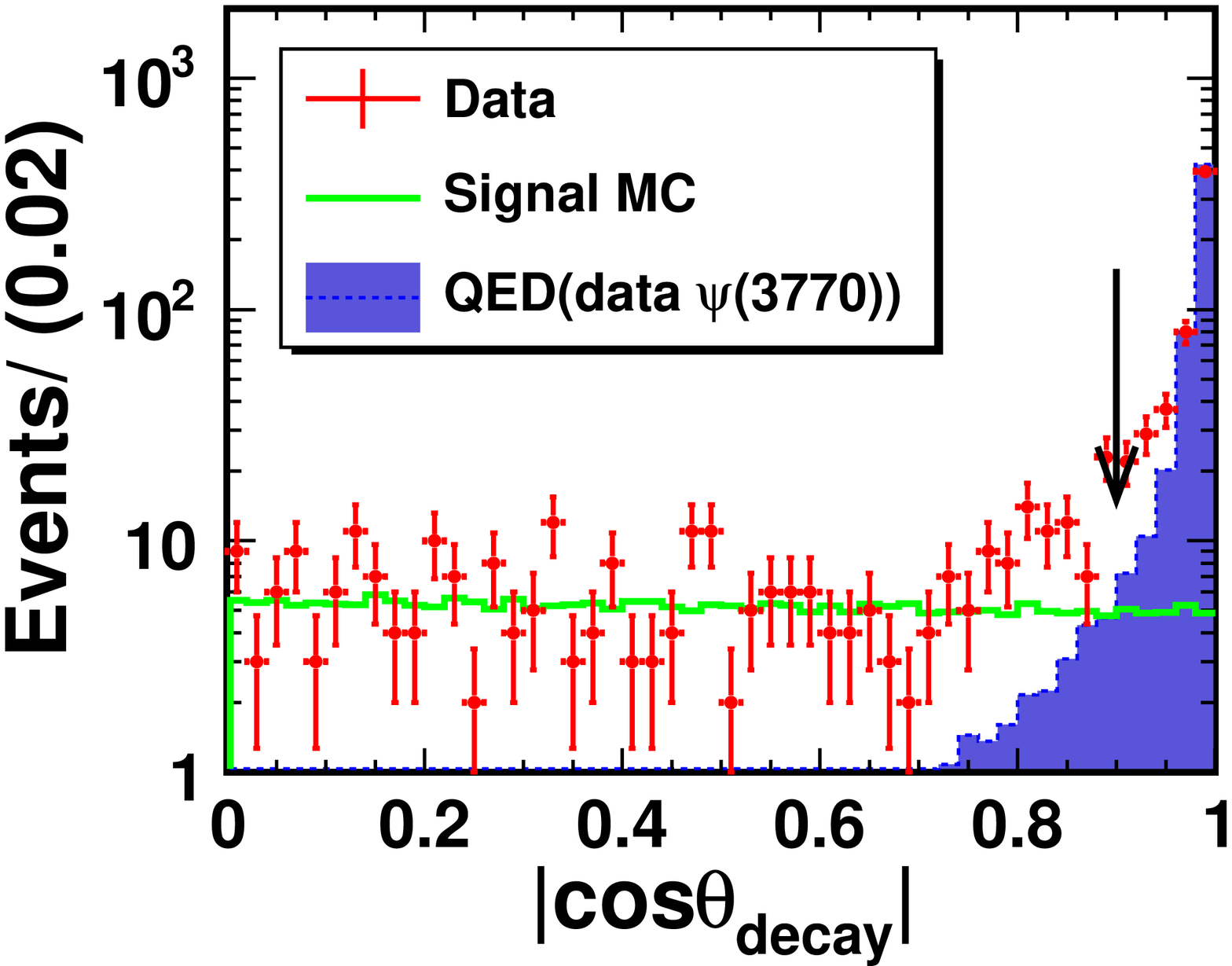}\put(-55,75){\bf \large~(a)}
        \includegraphics[width=4.4cm]{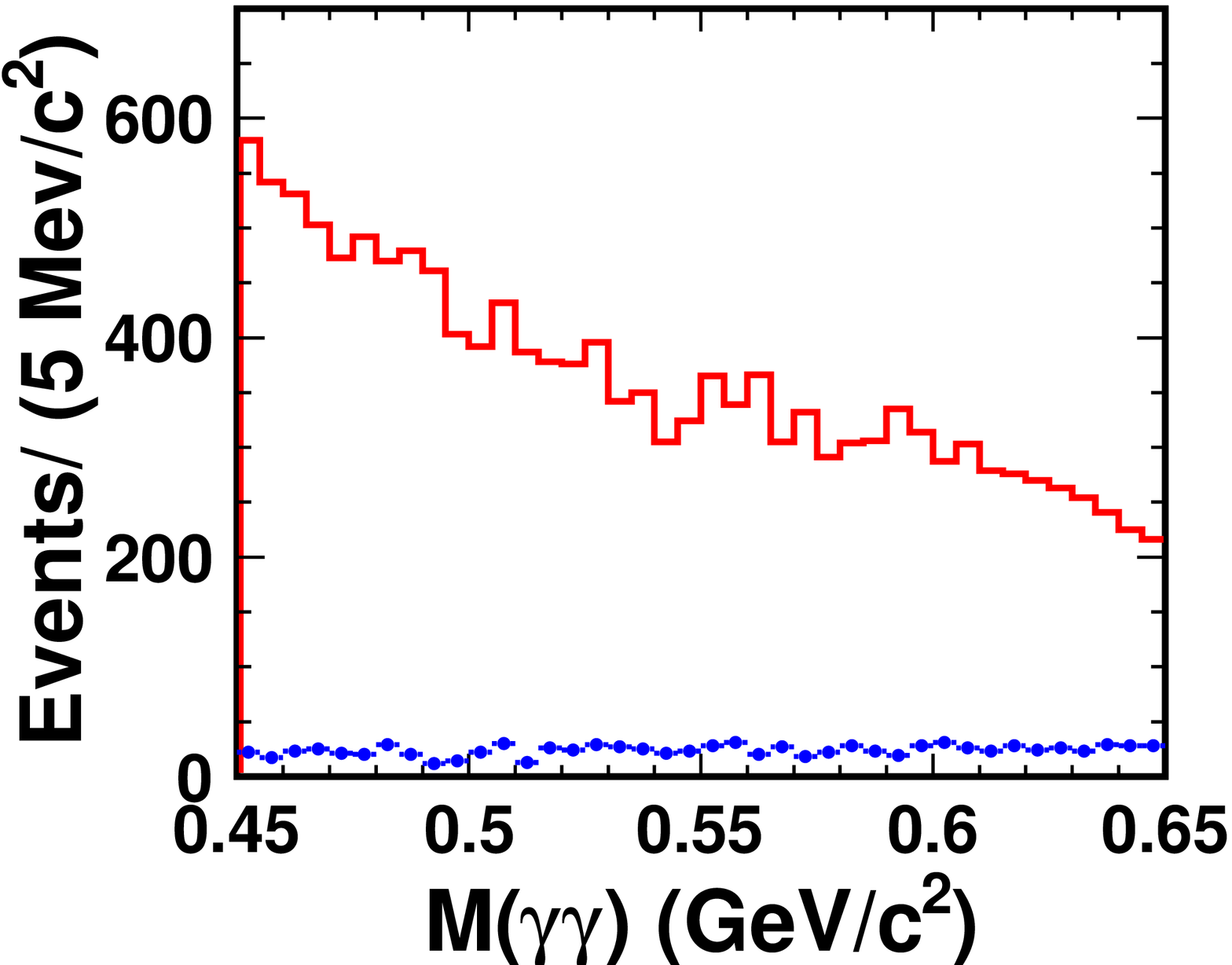}\put(-55,75){\bf \large~(b)}
        \\
        \includegraphics[width=4.4cm]{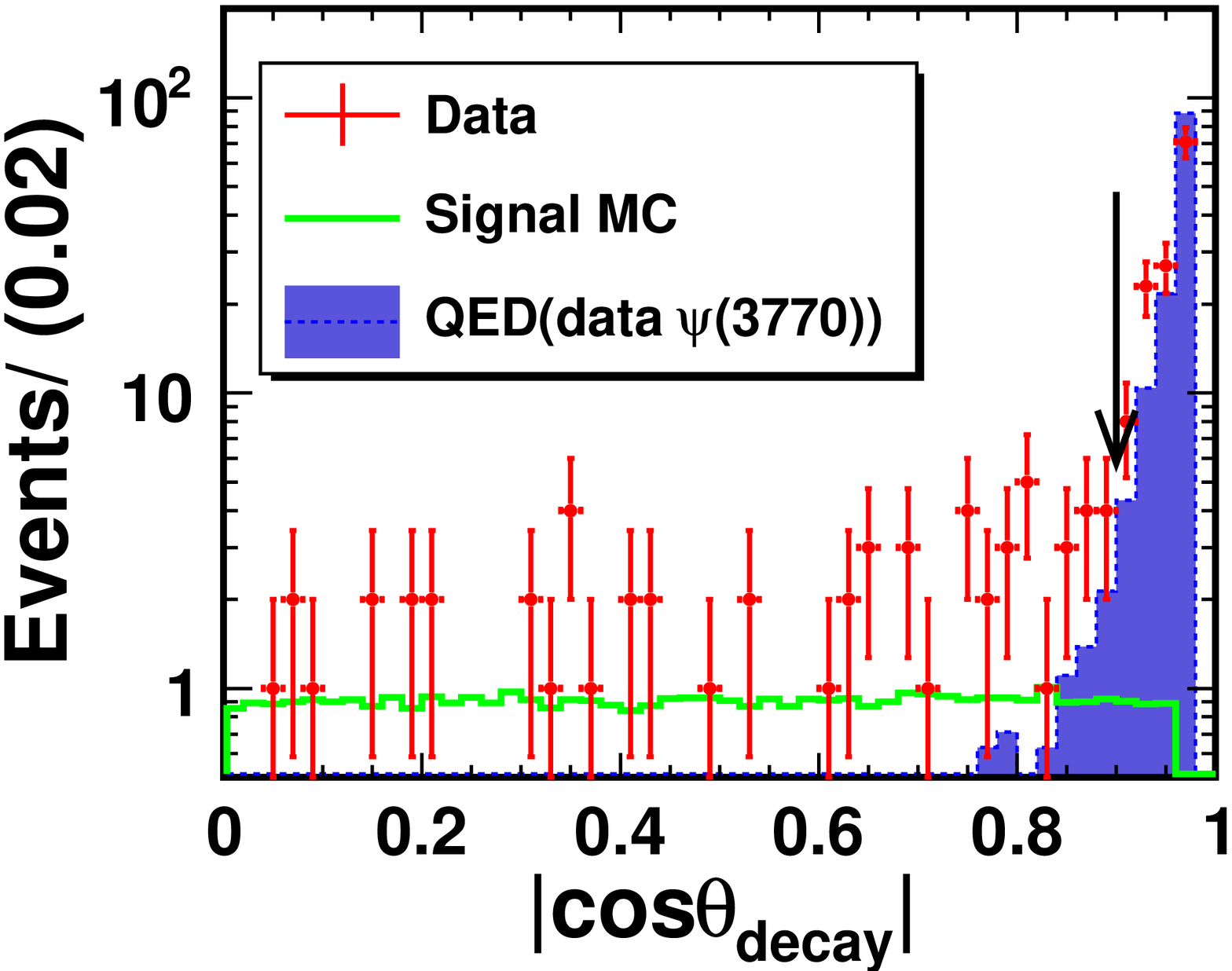}\put(-55,75){\bf \large~(c)}
        \includegraphics[width=4.4cm]{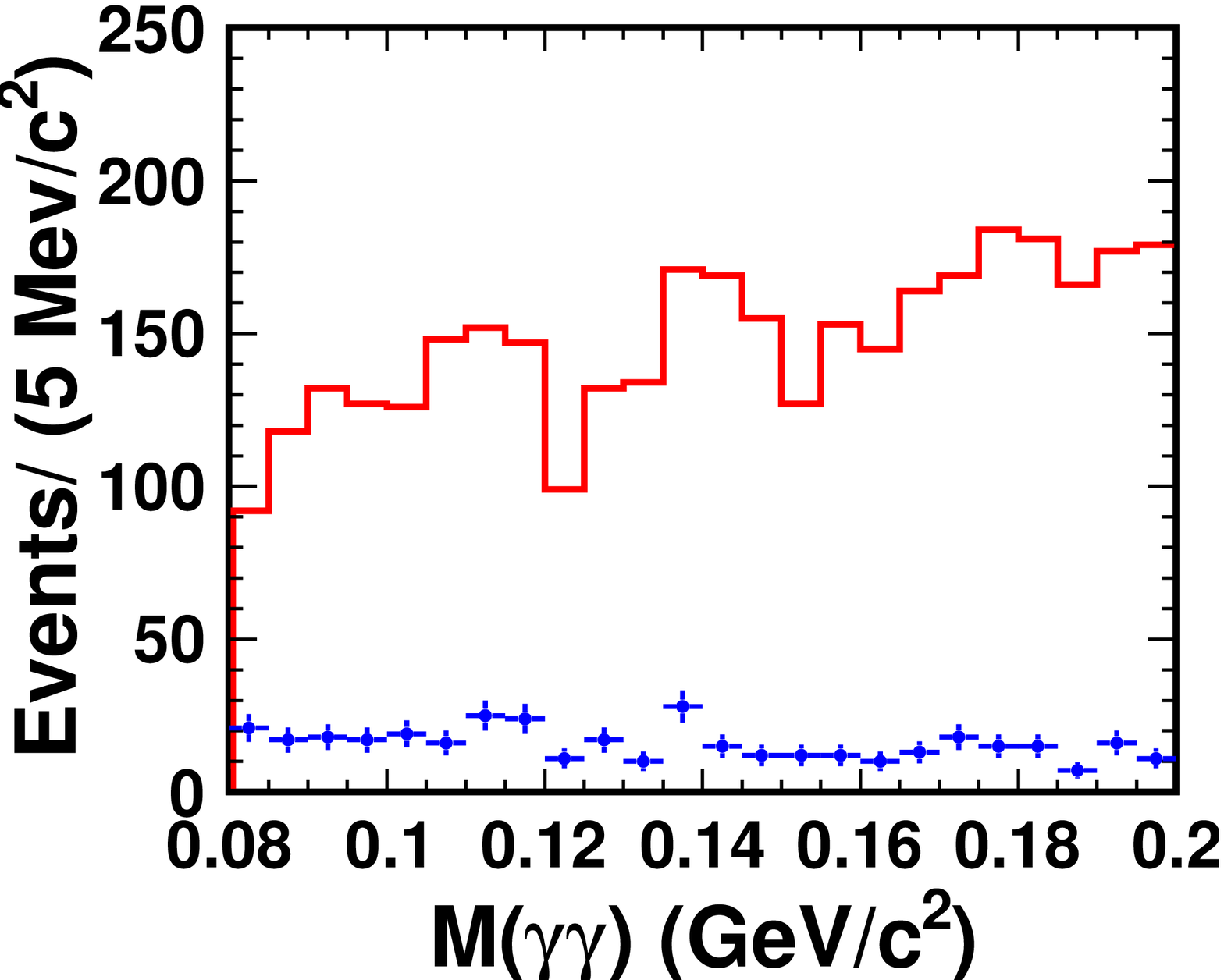}\put(-55,75){\bf \large~(d)}
        \caption{The $|\rm{cos}\theta_{\rm{decay}}|$  distributions
        (a) for $\eta$ and (c) for $\pi^0$, and two-photon invariant-mass
        distributions (b) for the $J/\psi \rightarrow
        \eta e^+ e^- (\eta \rightarrow \gamma \gamma)$ and (d) for the  $J/\psi \rightarrow \pi^0 e^+e^-
        (\pi^0 \rightarrow \gamma \gamma)$ modes. In (a) and(c), the (green) solid histograms are the MC-simulated signals, the (red) dots with error bars are
        data,
        the (blue) dotted histograms are from the $\psi(3770)$ data. The arrows indicate the requirement $|\rm{cos}\theta_{\rm{decay}}|<0.9$.
        In (b) and (d), the (red) histograms and the (blue) dots with error bars are $\psi(3770)$ data (used as a continuum sample) without and with the requirement, respectively.}
    \label{helicity-plots}
\end{figure}

For the $J/\psi \to \etap e^+e^- (\etap \to \gamma\pi^+\pi^-)$ and
$J/\psi \to \eta e^+e^- (\eta \to \pi^+\pi^-\pi^0)$ modes, there
are non-peaking backgrounds mainly coming from two sources.
One is from $J/\psi\to\gamma\pi^+\pi^-\pi^+\pi^-$ and
$J/\psi\to\pi^0\pi^+\pi^-\pi^+\pi^-$.
With two pions misidentified as an electron-positron pair,
this produces a smooth background under the $\etap$ or $\eta$ mass.
The other contribution is from
$J/\psi\to\pi^+ \pi^- \eta$, $\eta\to\gamma e^+ e^-$ and
$J/\psi\to\pi^+\pi^-\pi^0$, $\pi^0\to\gamma e^+ e^-$ with the same
final states as the signal mode $J/\psi \to \etap e^+e^- (\etap \to
\gamma\pi^+\pi^-)$.  The combined decay rate of $J/\psi\to\pi^+ \pi^- \eta$,
$\eta\to\gamma e^+ e^-$ is at the rate of $10^{-6}$; the net contribution is negligible
according to the MC simulations.  In order
to reject background from $J/\psi\to\pi^+\pi^-\pi^0 (\pi^0 \to \gamma e^+
e^-)$, we veto candidates with an invariant $\gamma e^+ e^-$ mass in the
interval $[0.10,0.16]$~\GeV; the remaining background contributes
a smooth shape under the $\etap$ mass.

For the $J/\psi \to \eta e^+e^-(\eta \to \gamma\gamma)$ and $J/\psi
\to \pi^0 e^+e^-(\pi^0 \to \gamma\gamma)$ modes, non-peaking continuum
backgrounds from the QED processes $e^+e^-\to e^+e^-\gamma(\gamma)$ and
$e^+e^-\to 3\gamma$ (in which one $\gamma$ converts into an $e^+e^-$
pair) are studied. Since $\eta$ and $\pi^0$ mesons decay
isotropically, the angular distribution of photons from $\eta$ or
$\pi^0$ decays is flat in $\theta_{\rm{decay}}$, the angle of the
decay photon in the $\eta$ or $\pi^0$ helicity frame.
However, continuum background events accumulate near
$\rm{cos}\theta_{\rm{decay}}=\pm 1$, and thus we require
$|\rm{cos}\theta_{\rm{decay}}|<0.9$. Figures~\ref{helicity-plots}
(a) and (c) show the $|\rm{cos}\theta_{\rm{decay}}|$ distributions
for $\eta$ and $\pi^0$ decays, respectively. The (blue) dotted
histogram peaking near $|\rm{cos}\theta_{\rm{decay}}| = 1$ in
Fig.~\ref{helicity-plots}(a) or (c) is from a 2.9~fb$^{-1}$
$\psi(3770)$ data sample taken at $\sqrt{s}=3.773$~GeV, which
is dominated by QED processes.
The MC events of $e^+e^-\to e^+e^-\gamma(\gamma)$
and $e^+e^-\to 3\gamma$ are generated using the Babayaga QED event
generator~\cite{babayaga} and the distributions are consistent with
that from the 3.773 GeV sample. After requiring $|\rm{cos}\theta_{\rm{decay}}|<0.9$,
as shown in Fig.~\ref{helicity-plots}(b) or (d), the background from QED
processes is reduced drastically.

\begin{figure}[htbp]
    \centering
        \includegraphics[width=4.5cm]{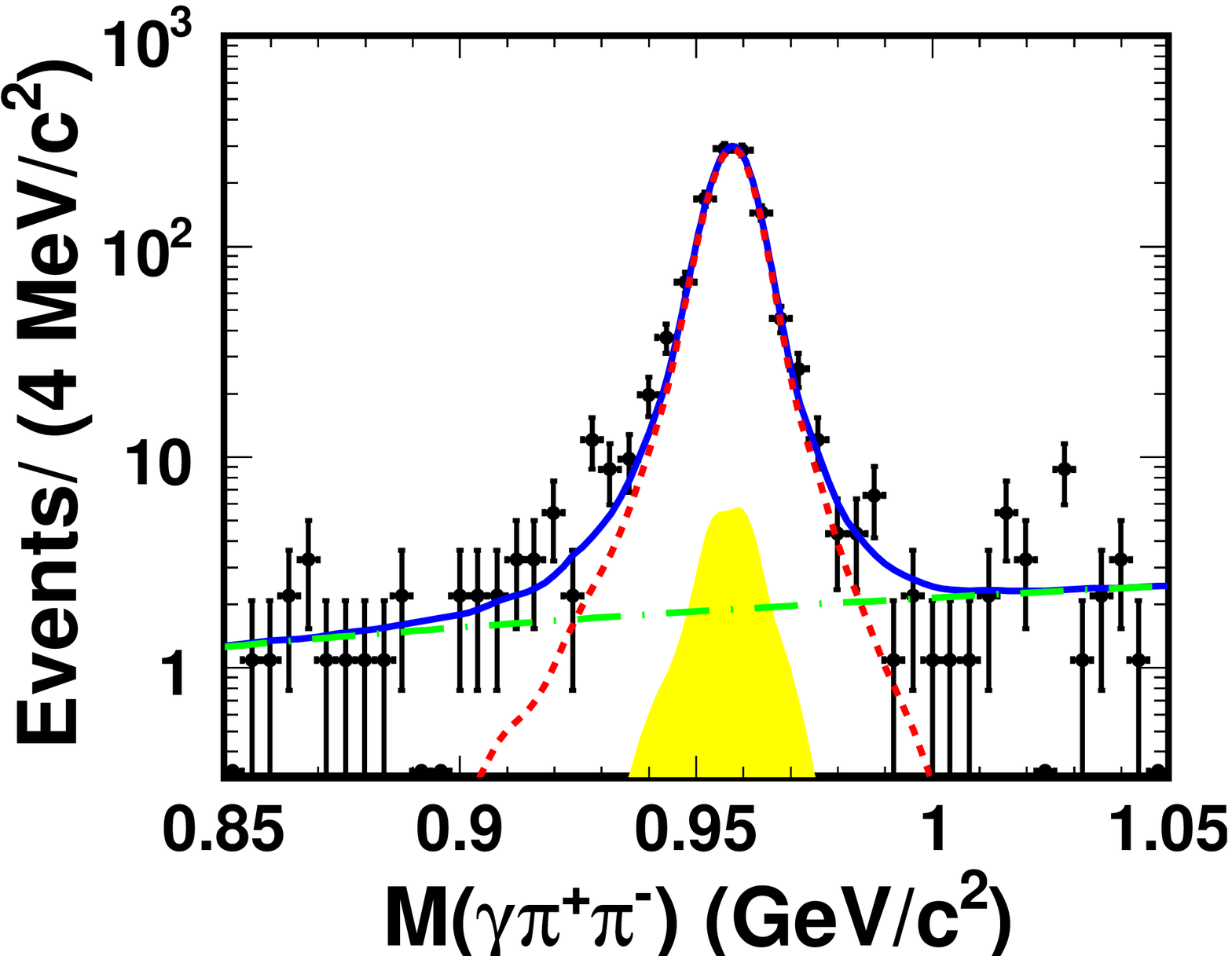}\put(-105,70){\bf \large~(a)}
        \includegraphics[width=4.5cm]{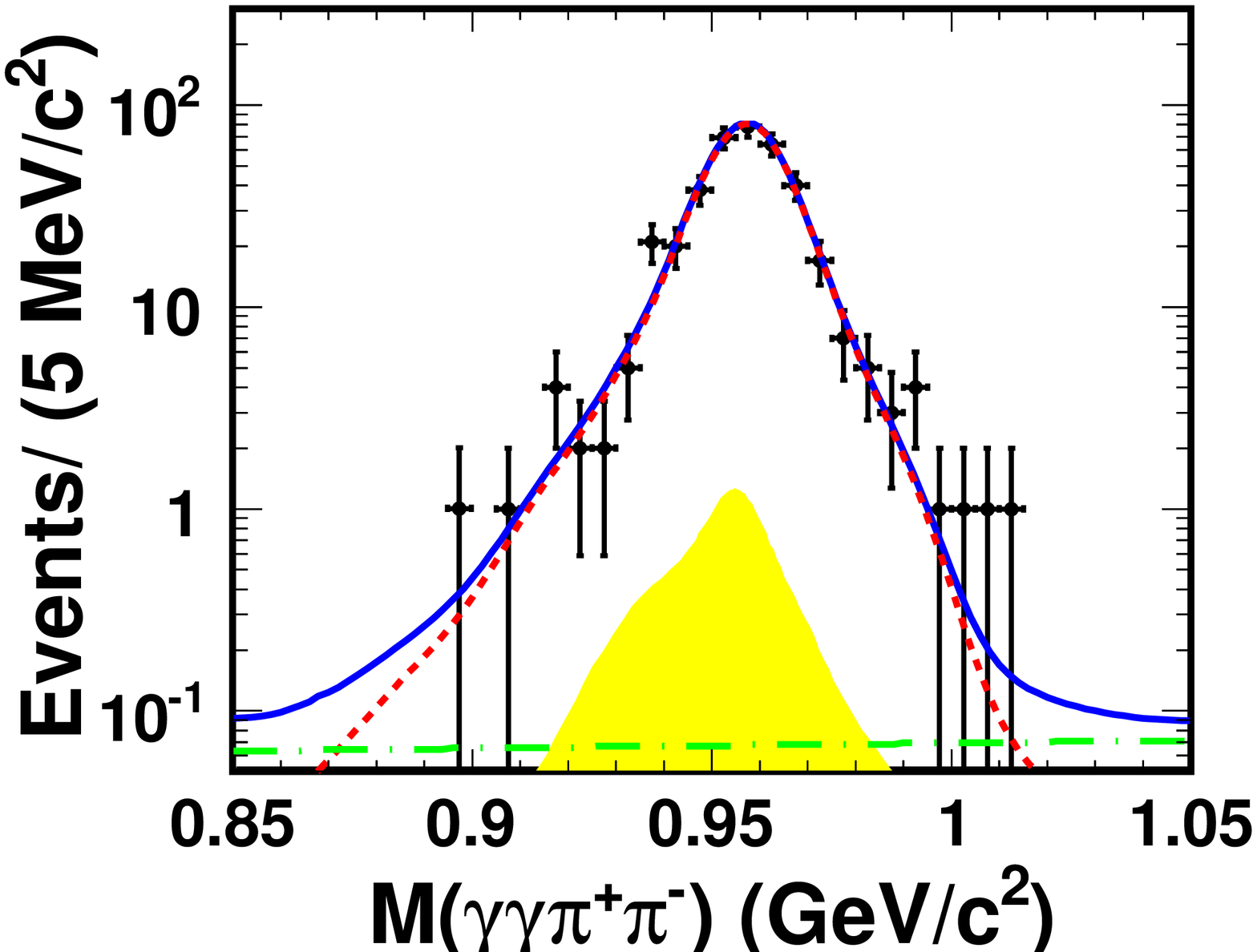}\put(-105,70){\bf \large~(b)}
         \\
        \includegraphics[width=4.5cm]{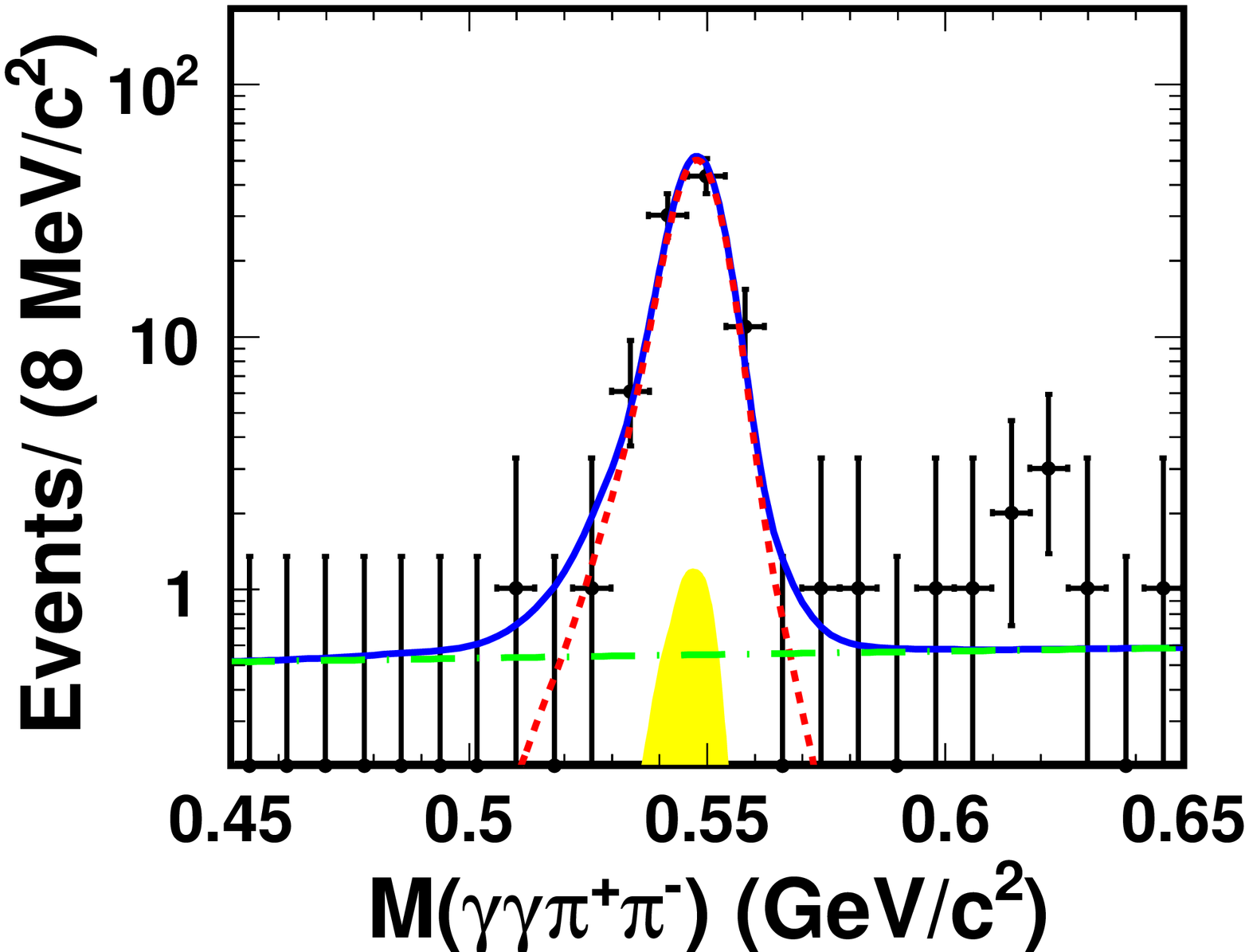}\put(-105,70){\bf \large~(c)}
        \includegraphics[width=4.5cm]{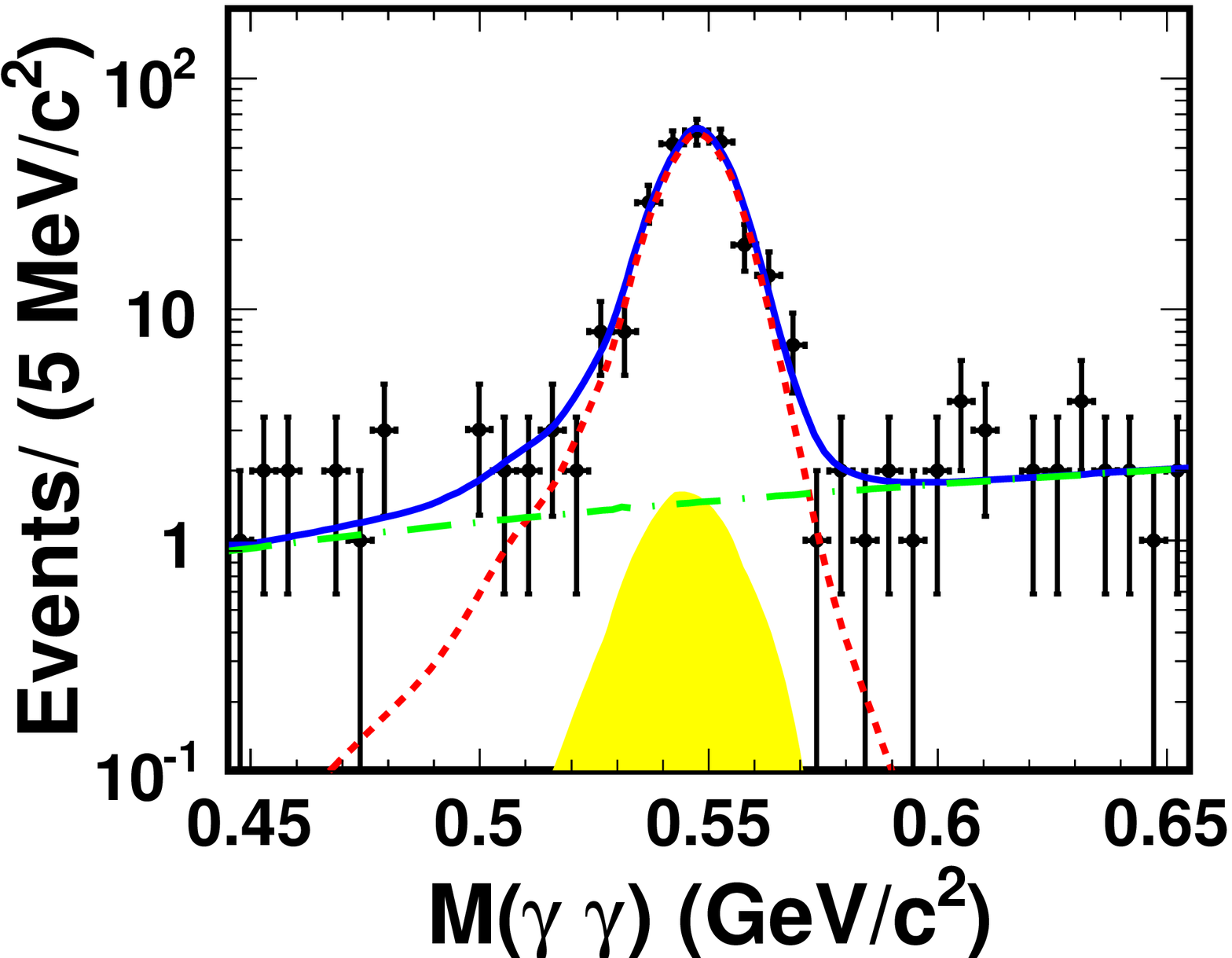}\put(-105,70){\bf \large~(d)}
         \\
          \hspace{-4.2cm}
        \includegraphics[width=4.5cm]{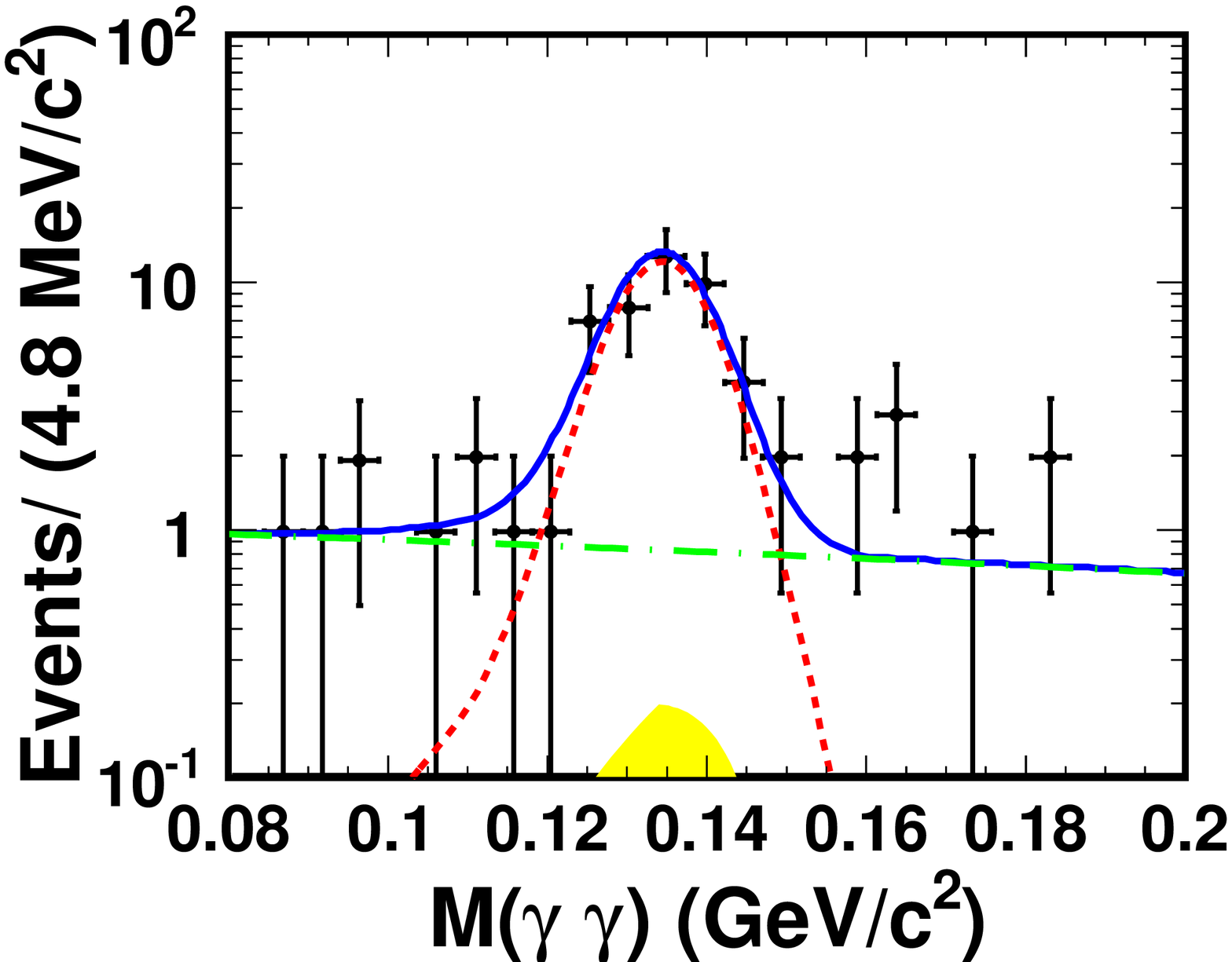}\put(-105,70){\bf \large~(e)}
        \caption{Mass distributions of the pseudoscalar
meson candidates in $J/\psi \rightarrow P e^+e^-$: (a) $\etap
\rightarrow \gamma \pi^+\pi^-$, (b) $\etap \rightarrow \pi^+\pi^-
\eta$ ($\eta \rightarrow \gamma \gamma$), (c) $\eta \rightarrow
\pi^+\pi^-\pi^0$, (d) $\eta \rightarrow \gamma \gamma$, and (e)
$\pi^0 \rightarrow \gamma \gamma$. The (black) dots with error bars are data, the
(red) dashed lines represent the signal, the (green) dot-dashed
curves shows the non-peaking background shapes, the (yellow) shaded
components are the shapes of the peaking backgrounds from the
$J/\psi \rightarrow P\gamma$ decays. Total fits are shown as the
(blue) solid lines. }
    \label{spectrum}
\end{figure}

Mass spectra of the signal modes with all of the selection criteria
applied are presented in Fig.~\ref{spectrum}.
The signal efficiencies determined from MC simulations for the $\etap$,
$\eta$ and $\pi^0$ are shown in Table~\ref{yields}.

An unbinned extended maximum likelihood (ML) fit is performed for
each mode to determine the event yield. The signal probability
density function (PDF) in each mode is represented by the signal MC
shape convoluted with a Gaussian function, with parameters
determined from the fit to the data. The Gaussian function is to
describe the MC-data difference due to resolution.  The shape for
the non-peaking background is described by a first- or second-order
Chebychev polynomial, and the background yield and its PDF
parameters are allowed to float in the fit. The dominant peaking
background from the $\gamma$-conversion events in the $J/\psi \to
P\gamma$ decay is obtained from the MC-simulated shape with the
number fixed to the normalized value. The fitting ranges for the
$\etap$, $\eta$ and $\pi^0$ modes are $0.85-1.05$~\GeV,
$0.45-0.65$~\GeV~ and $0.08-0.20$~\GeV, respectively. As discussed
in Section~\ref{Analysis}, the estimated numbers of peaking
background events are subtracted from the fitted yields. The net
signal yields for all modes are summarized in Table~\ref{yields}.

\begin{figure}[htbp]
    \centering
        \includegraphics[width=4.5cm]{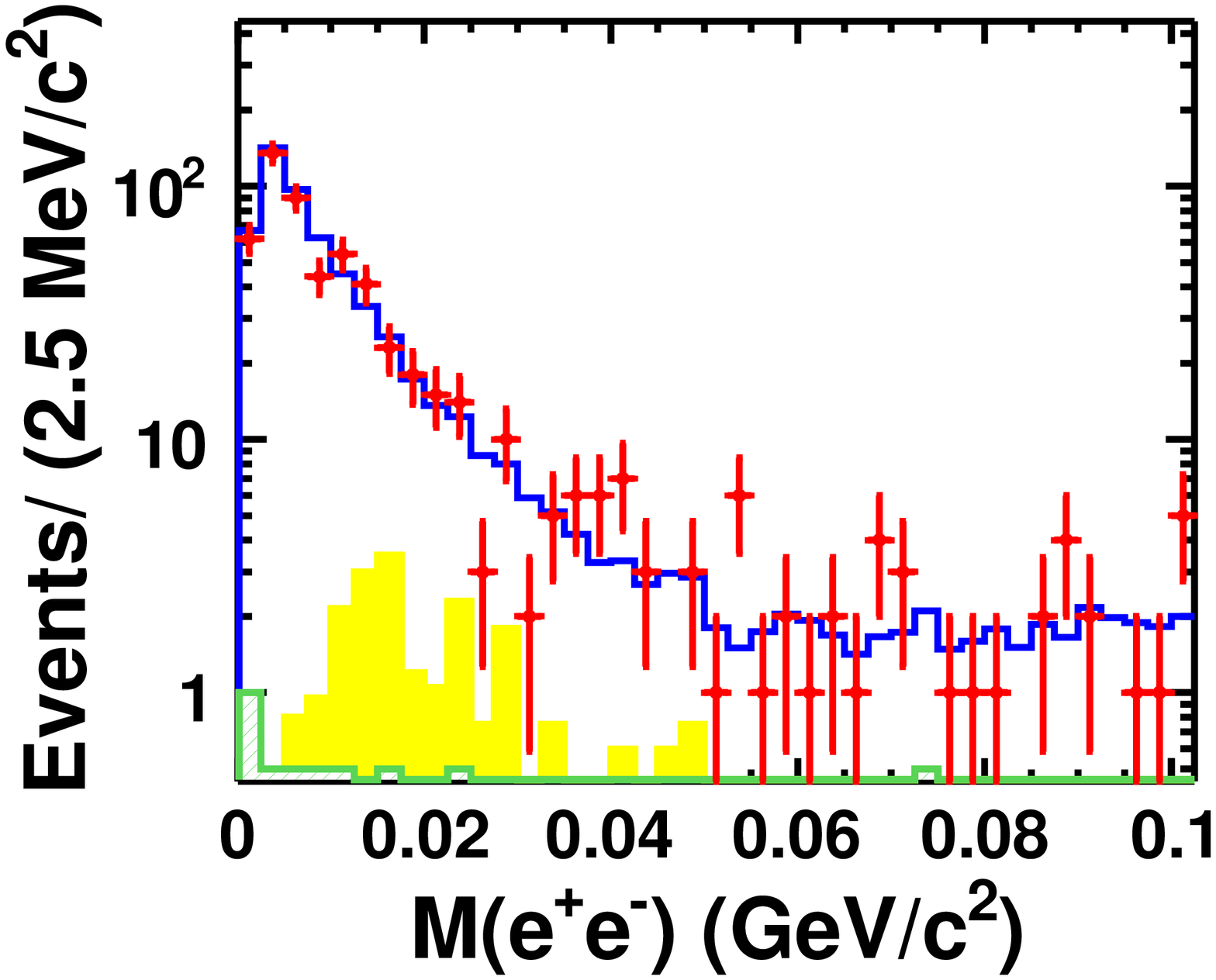}\put(-40,65){\bf \large~(a)}
        \includegraphics[width=4.5cm]{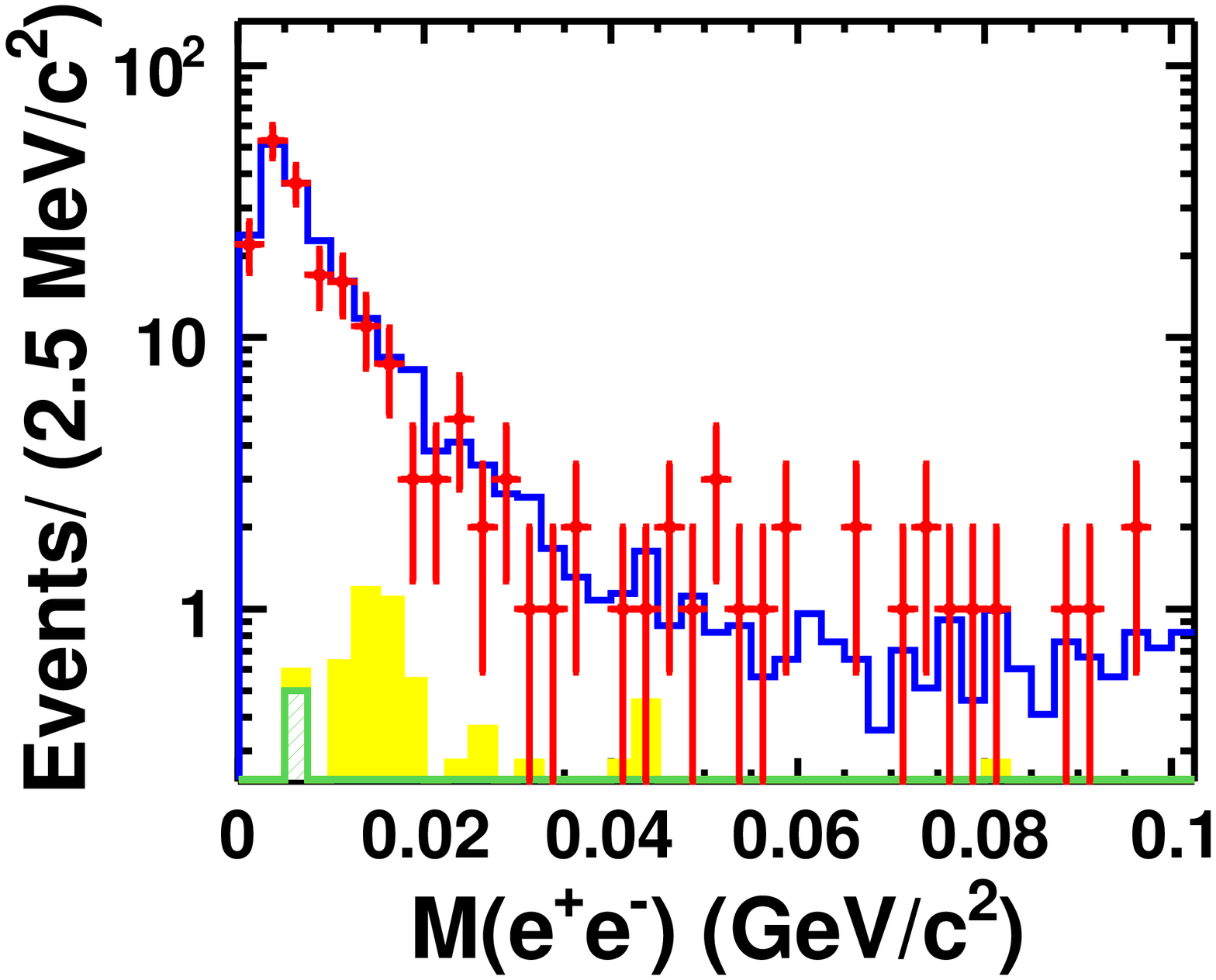}\put(-40,65){\bf \large~(b)}
         \\
        \includegraphics[width=4.5cm]{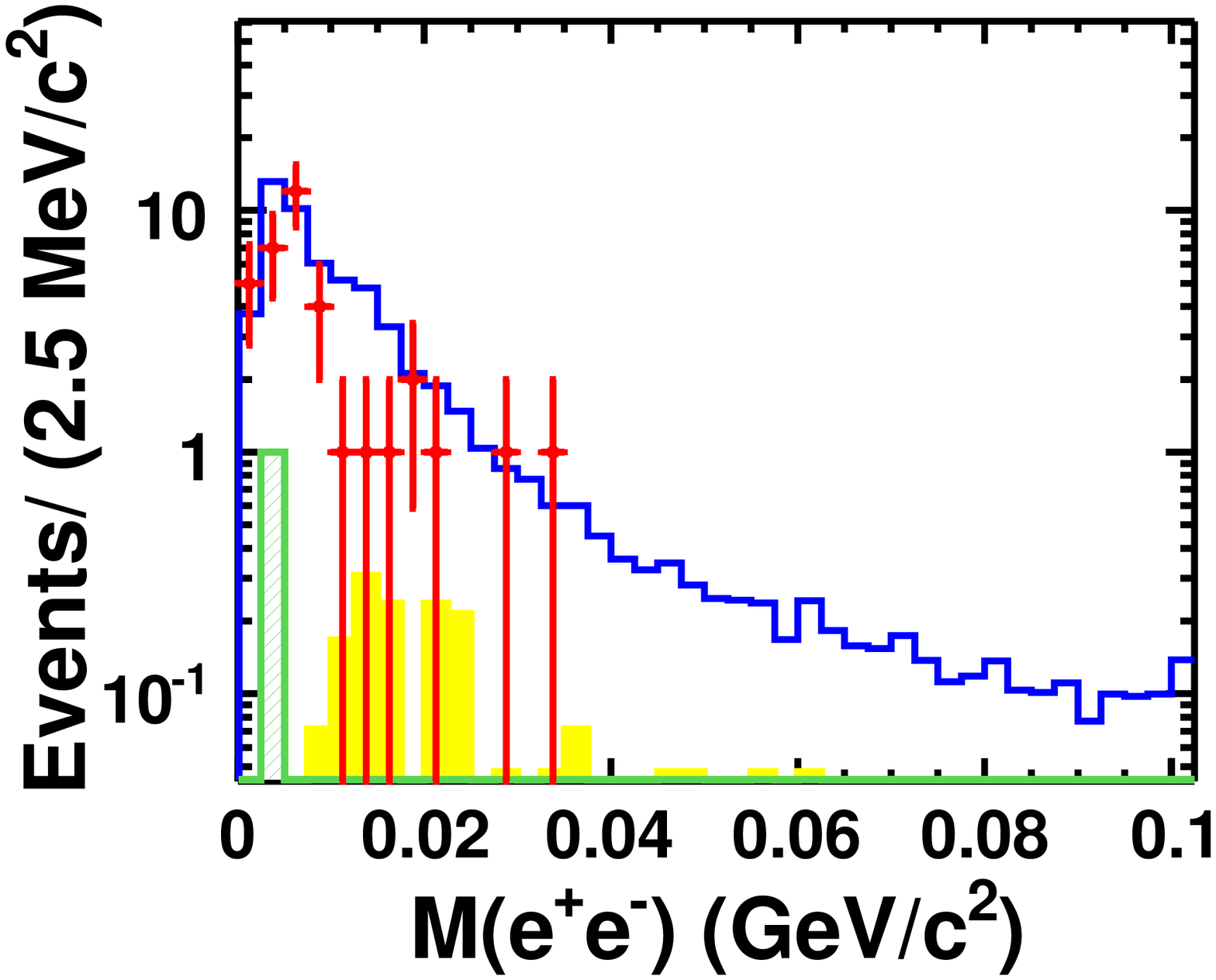}\put(-40,65){\bf \large~(c)}
        \includegraphics[width=4.5cm]{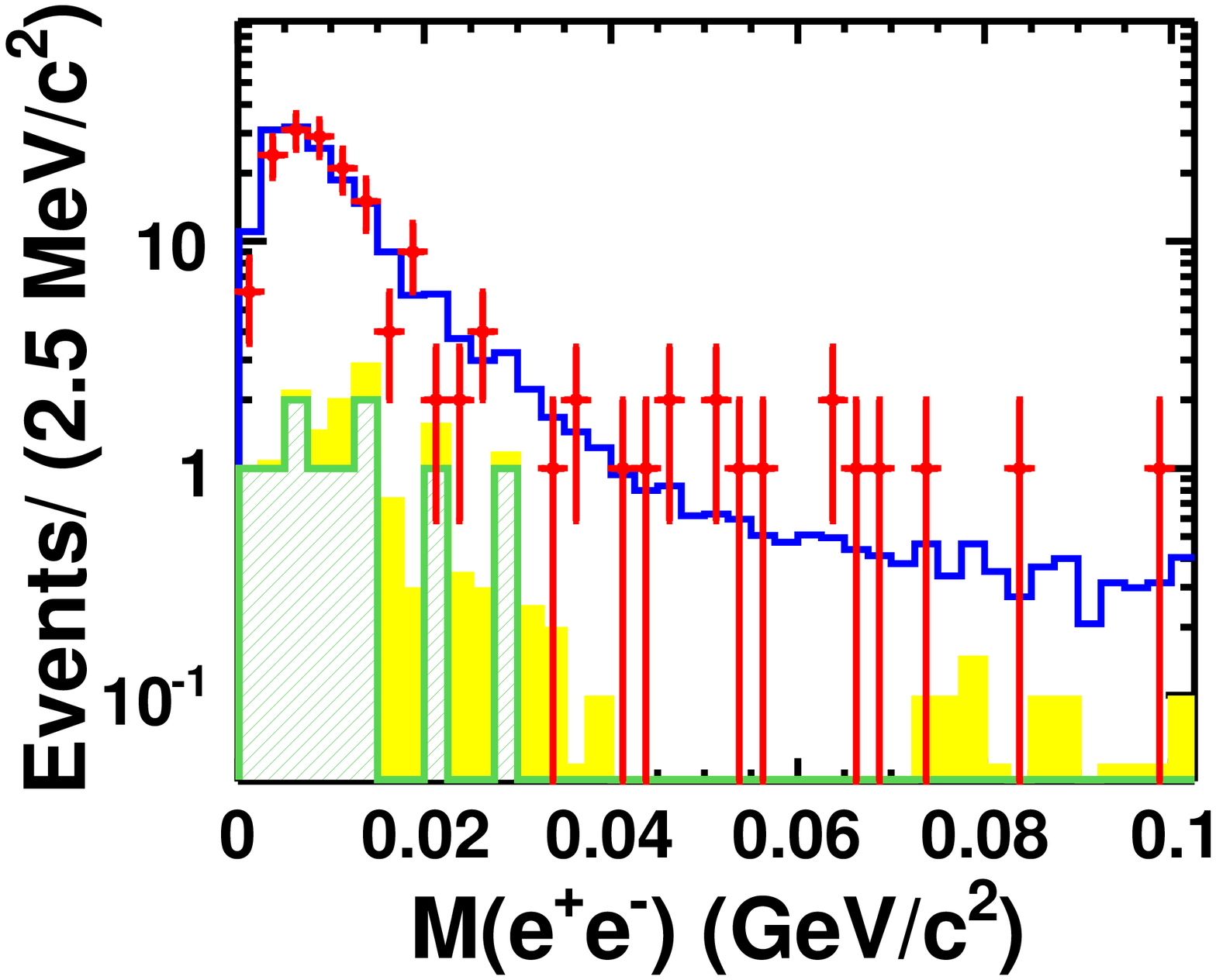}\put(-40,65){\bf \large~(d)}
         \\
          \hspace{-4.2cm}
        \includegraphics[width=4.5cm]{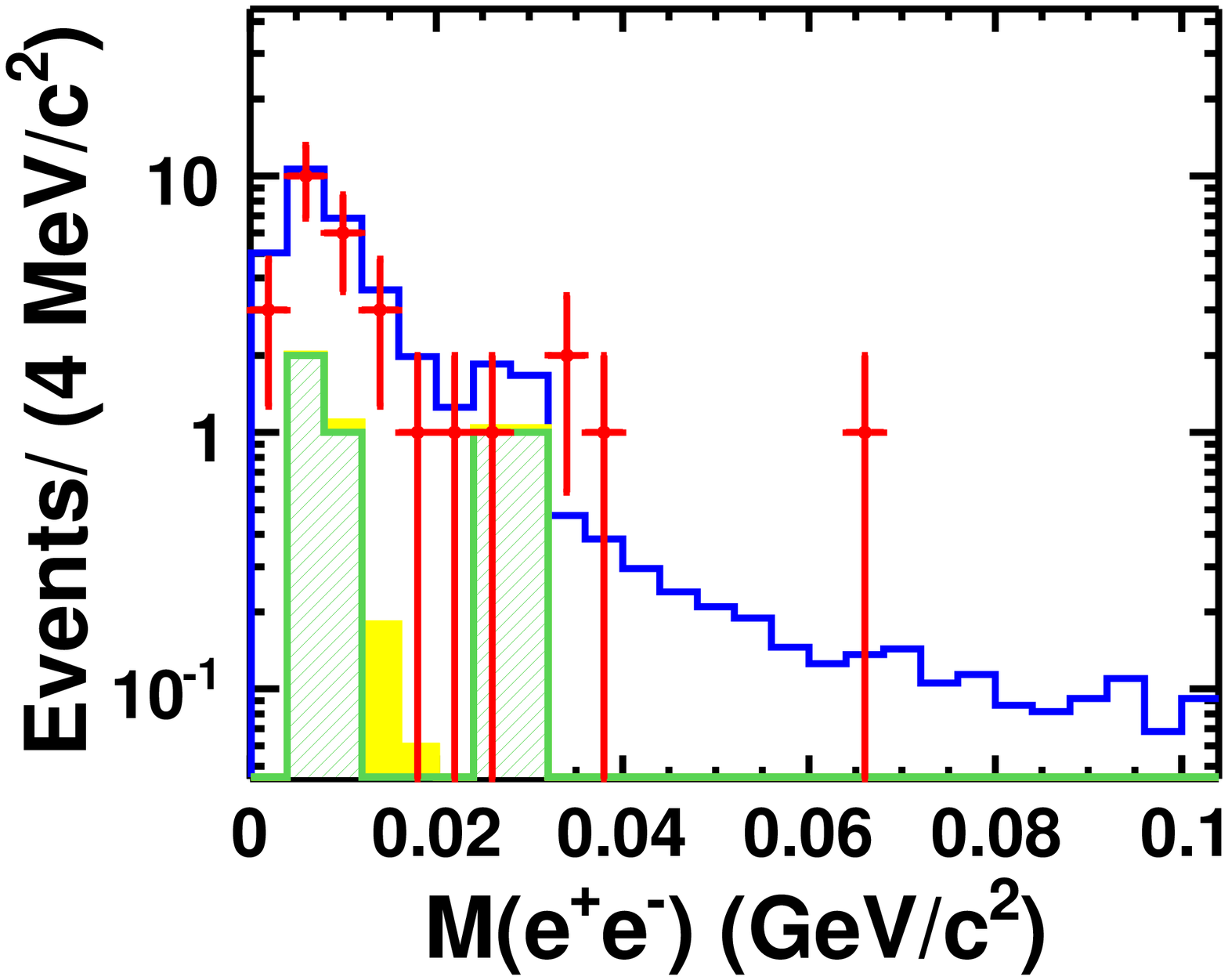}\put(-40,65){\bf \large~(e)}
        \caption{The $M_{e^+e^-}$ mass distributions in $J/\psi \rightarrow P e^+e^-$: (a) $\etap
\rightarrow \gamma \pi^+\pi^-$, (b) $\etap \rightarrow \pi^+\pi^-
\eta$ ($\eta \rightarrow \gamma \gamma$), (c) $\eta \rightarrow
\pi^+\pi^-\pi^0$, (d) $\eta \rightarrow \gamma \gamma$, and (e)
$\pi^0 \rightarrow \gamma \gamma$. The (red) dots with error bars are data, the (yellow)
shaded components are from the $\gamma$-conversion backgrounds in
the $J/\psi \rightarrow P\gamma$ decays, the (green) light-shaded
histograms are from non-peaking backgrounds estimated from the
sidebands on the pseudoscalar mass spectra. The (blue) histograms represent the sum of backgrounds and MC-simulated signals.}
    \label{Mee}
\end{figure}

To further demonstrate the high quality of signal events, the candidate events
within  $\pm 3\sigma$ of the pseudoscalar meson mass region for
each mode are projected to the $M_{e^+e^-}$ mass distribution in the
region of $[0.0, 0.1]$~\GeV~ as shown in Fig.~\ref{Mee}.
The signal MC events are generated based on the amplitude squared in Eq.(3)
of Ref.~\cite{Jinlin} for each mode, normalized to the fitted yield.
The number of the peaking backgrounds from $\gamma$-conversion events
is fixed to the expected value, and the non-peaking backgrounds
are estimated by using the sidebands of the pseudoscalar mass spectra.
The consistency of the data shapes with signal MC events indicates clear
signals in all modes for the EM Dalitz decays $J/\psi \rightarrow Pe^+e^-$.

\section{Systematic uncertainties}

Table~\ref{sum:sys} compiles all sources of systematic
uncertainties in the measurement of the branching fractions. Most
systematic uncertainties are determined from
comparisons of clean, high statistics test samples with results
from MC simulations.

The MDC tracking efficiency of the charged pion  is studied using
the control samples of $\psip \to \pi^+\pi^- J/\psi$, $J/\psi
\rightarrow l^+l^-$ ($l=e$, $\mu$) and $J/\psi \rightarrow
\pi^+\pi^-\pi^0$~\cite{photon}. The difference between data and MC
simulation is 1.0\% for each charged pion. The tracking efficiency
for the electron or positron is obtained with the control sample of
radiative Bhabha scattering $e^+e^-\rightarrow \gamma e^+e^-$
(including $J/\psi \to \gamma e^+ e^-$) at the $J/\psi$ energy
point. The tracking efficiency is calculated with
$\epsilon_{\rm{electron}} = N_{\rm{full}}/N_{\rm{all}}$, where
$N_{\rm{full}}$ indicates the number of $\gamma e^+e^-$ events with
all final tracks reconstructed successfully; and $N_{\rm{all}}$
indicates the number of events with one or both charged lepton
particles successfully reconstructed in addition to the radiative
photon. The difference in tracking efficiency between data and MC
simulation is calculated bin-by-bin over the distribution of
transverse momentum versus the polar angle of the lepton tracks.
The uncertainty is determined to be 1.0\%  per electron.
Tracking uncertainties are treated as fully correlated and thus
added linearly.

The photon detection efficiency and its uncertainty
 are studied using three different methods
described in Ref.~\cite{photon}. On average, the efficiency
difference between data and MC simulation is less than 1.0\% per
photon~\cite{photon}. The uncertainty from $\pi^0$ reconstruction is
determined to be 1.0\% per $\pi^0$ from the control sample
$J/\psi \to \pi^+\pi^-\pi^0$~\cite{pi0Rec}, and that for
$\eta$ reconstruction is 1.0\% from the control sample
$J/\psi \to p\bar{p}\eta$~\cite{pi0Rec}.

 The uncertainty on electron identification is
studied with the control sample of radiative Bhabha scattering
$e^+e^- \rightarrow \gamma e^+e^-$ (including $J/\psi \rightarrow
\gamma e^+e^-$); samples with backgrounds less than 1.0\% are
obtained~\cite{zhush}. The efficiency difference for electron identification
between the data and MC simulation of about 1.0\% is taken as our uncertainty.

\begin{table*}[hbtp]
\begin{center}
\vspace{-0.0cm} \caption{ {{Summary of systematic
uncertainties (\%). The terms with asterisks are correlated systematic
uncertainties between $\etap\to\gamma\pi^+\pi^-$ and
$\etap\to\pi^+\pi^-\eta$ ($\eta\to\pi^+\pi^-\pi^0$ and $\eta\to
\gamma\gamma$).   }} }
  \label{sum:sys}
  \vspace{0.3cm}
    \renewcommand{\arraystretch}{1.5}
     \begin{tabular}{lccccc}
        \hline\hline

        ~&\hspace{0.3cm}$\etap\to\gamma\pi^+\pi^-$ \hspace{0.3cm}&\hspace{0.3cm}$\etap\to\pi^+\pi^-\eta$\hspace{0.3cm}& \hspace{0.3cm}$\eta\to\pi^+\pi^-\pi^0$ \hspace{0.3cm}& \hspace{0.3cm}$\eta\to \gamma\gamma$ \hspace{0.3cm}&\hspace{0.3cm} $\pi^0\to\gamma\gamma$ \hspace{0.3cm}\\ \hline

        MDC tracking$^*$                      & 4.0   & 4.0   &  4.0   &   2.0  &  2.0   \\

        Photon detection $^*$                 & 1.0   & 2.0    & 2.0   &  2.0  &  2.0  \\

        $\pi^0(\eta)$ reconstruction          & --   & 1.0    & 1.0   &  1.0  &  1.0   \\

        Electron identification$^*$           & 2.0  & 2.0    & 2.0   &  2.0   &  2.0  \\

        Veto of the $\gamma$-conversion$^*$    & 1.0  & 1.0   & 1.0    & 1.0    & 1.0   \\

        4C kinematic fit                  & 1.0  & 1.0    &1.0    &1.0     &1.0    \\

        Form factor                       & 1.0  & 1.1   & 1.1    & 2.2    & 3.1  \\

        Signal shape                      & 0.9  & 0.5    & 0.8    & 0.1    & 1.0 \\

        Background shape                  & 0.9  & 1.0   & 1.0  &  2.7   &  4.0  \\

        Cited branching fractions         & 2.0   & 1.7    &1.2   &  0.5  &  0.0  \\

        Number of $J/\psi$$^*$            & 1.2   & 1.2    &1.2   &  1.2  &  1.2  \\ \hline

        Total                             & 5.6   & 5.8   &5.7   &  5.4  &  6.6  \\
        \hline\hline
      \end{tabular}
      \vspace{-0.5cm}
\end{center}
\end{table*}

\begin{figure}[htbp]
    \centering
        \includegraphics[width=4.4cm]{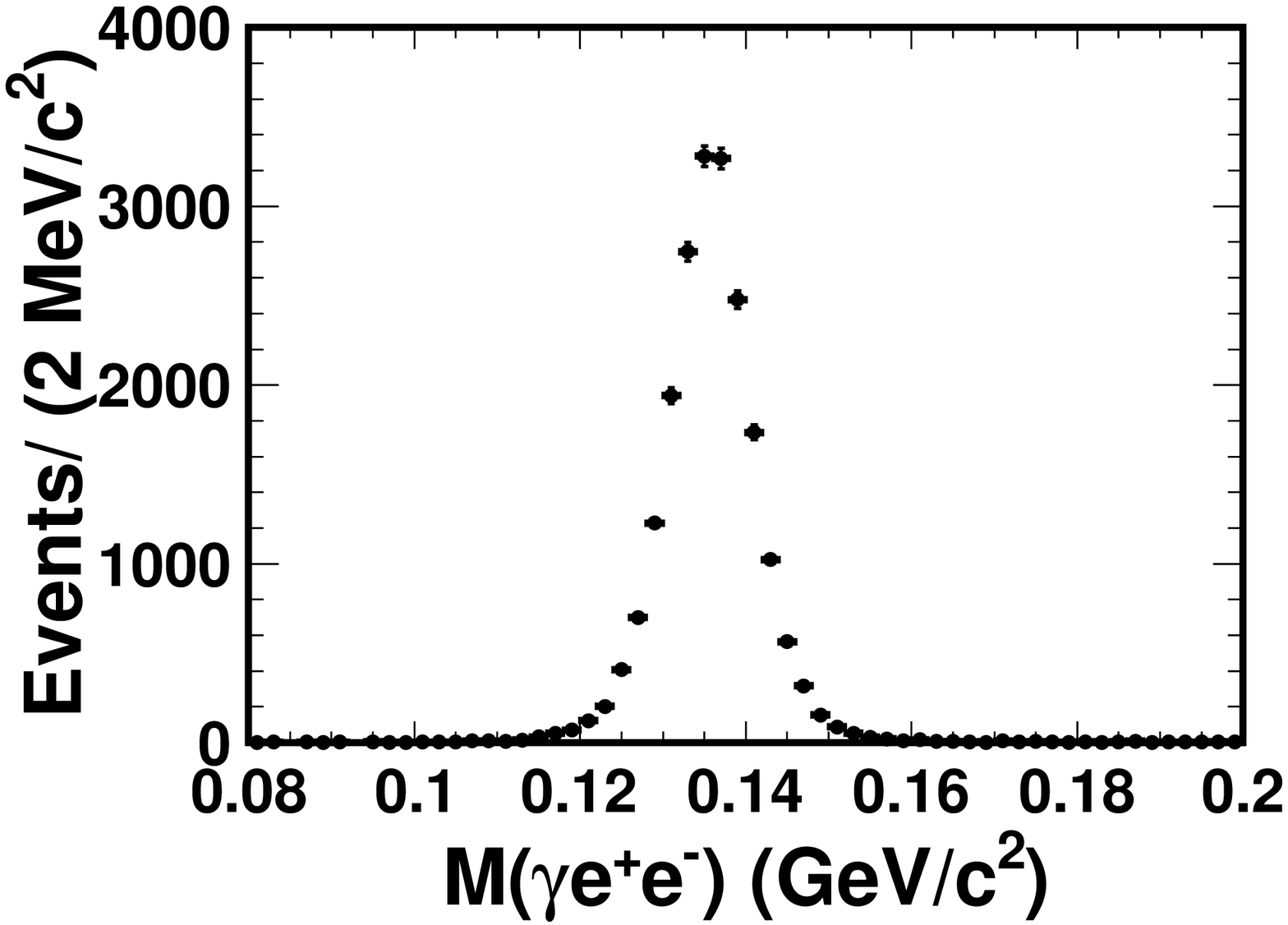}\put(-45,68){\bf \large~(a)}
        \includegraphics[width=4.4cm]{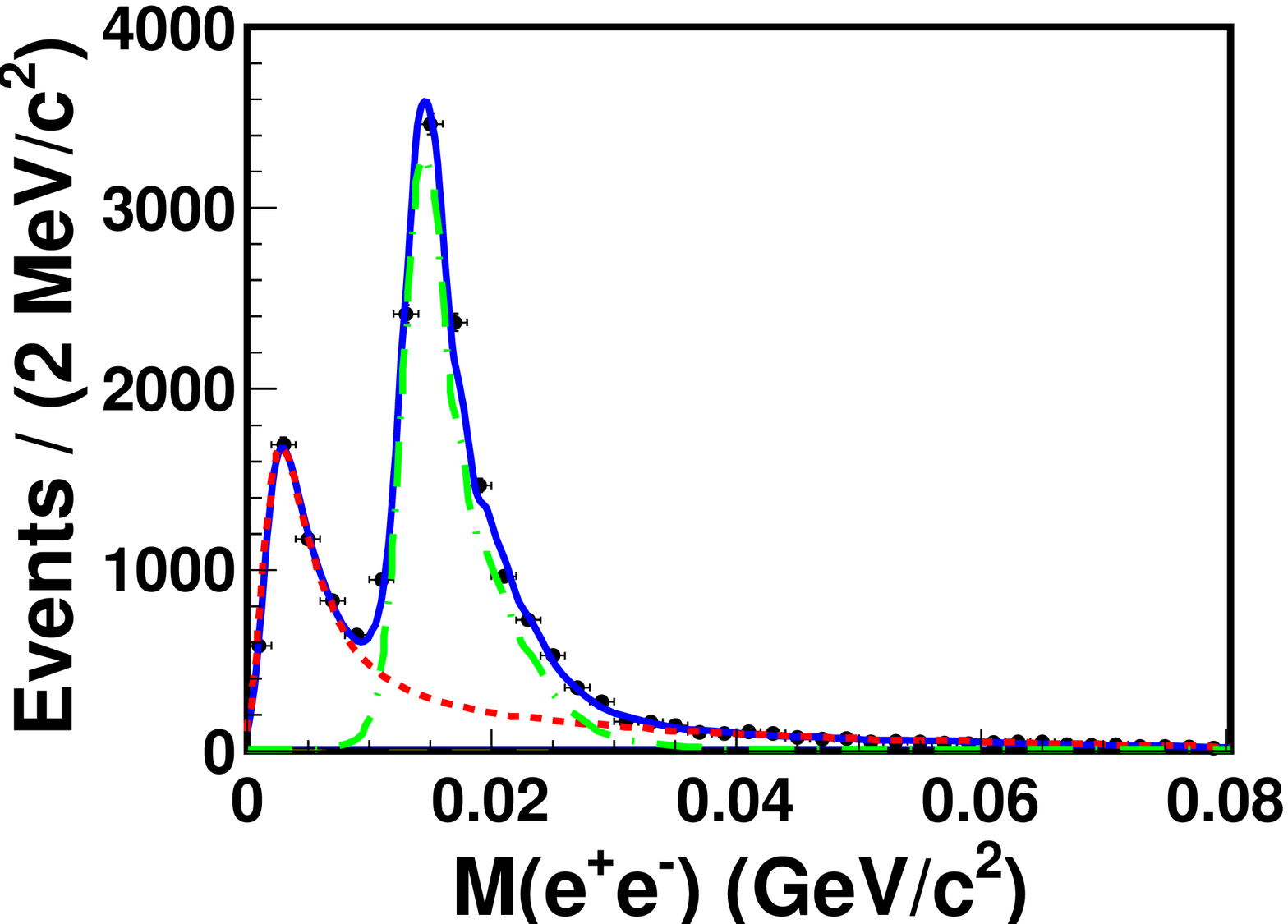}\put(-45,68){\bf \large~(b)}
        \\
        \includegraphics[width=4.4cm]{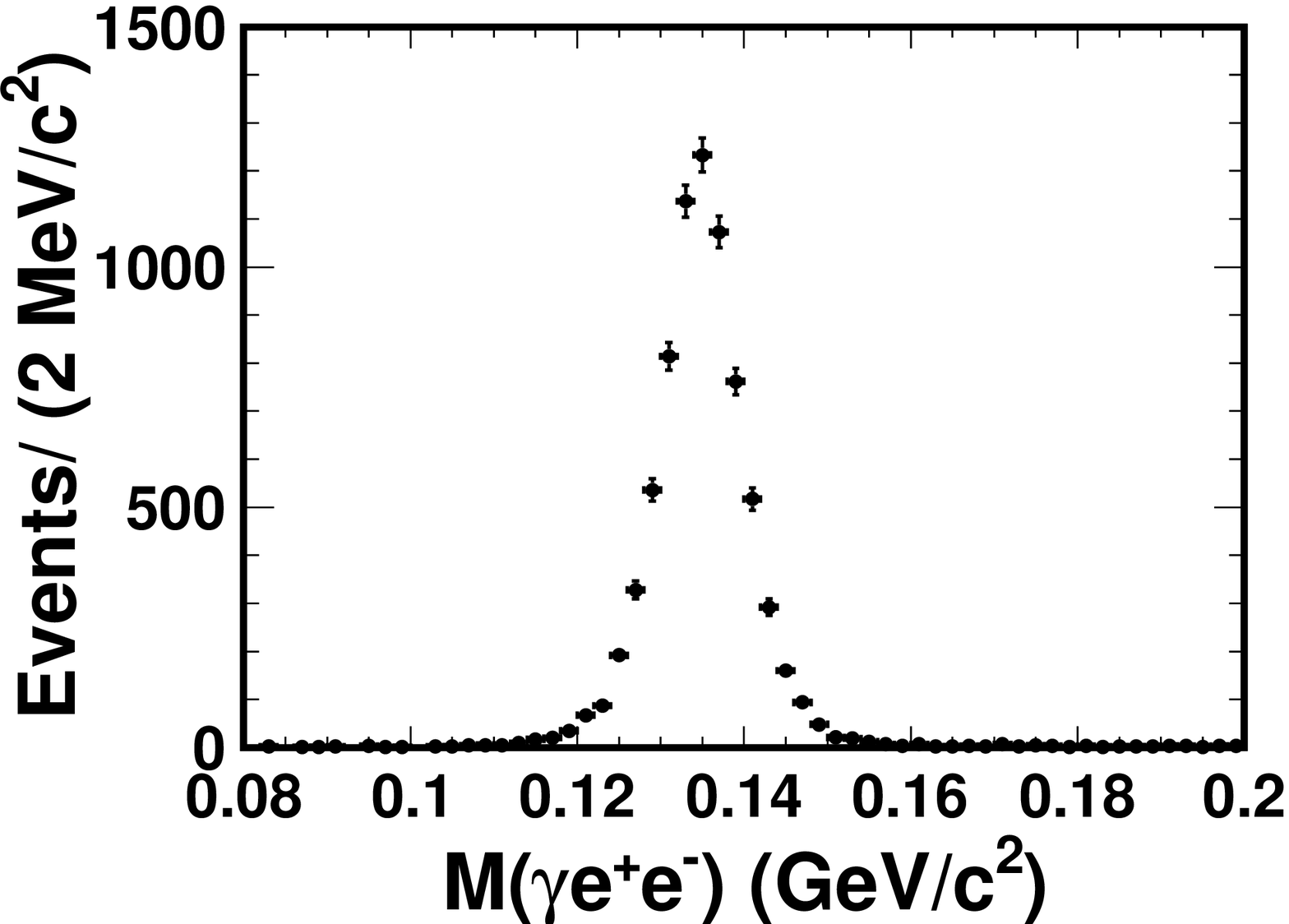}\put(-45,68){\bf \large~(c)}
        \includegraphics[width=4.4cm]{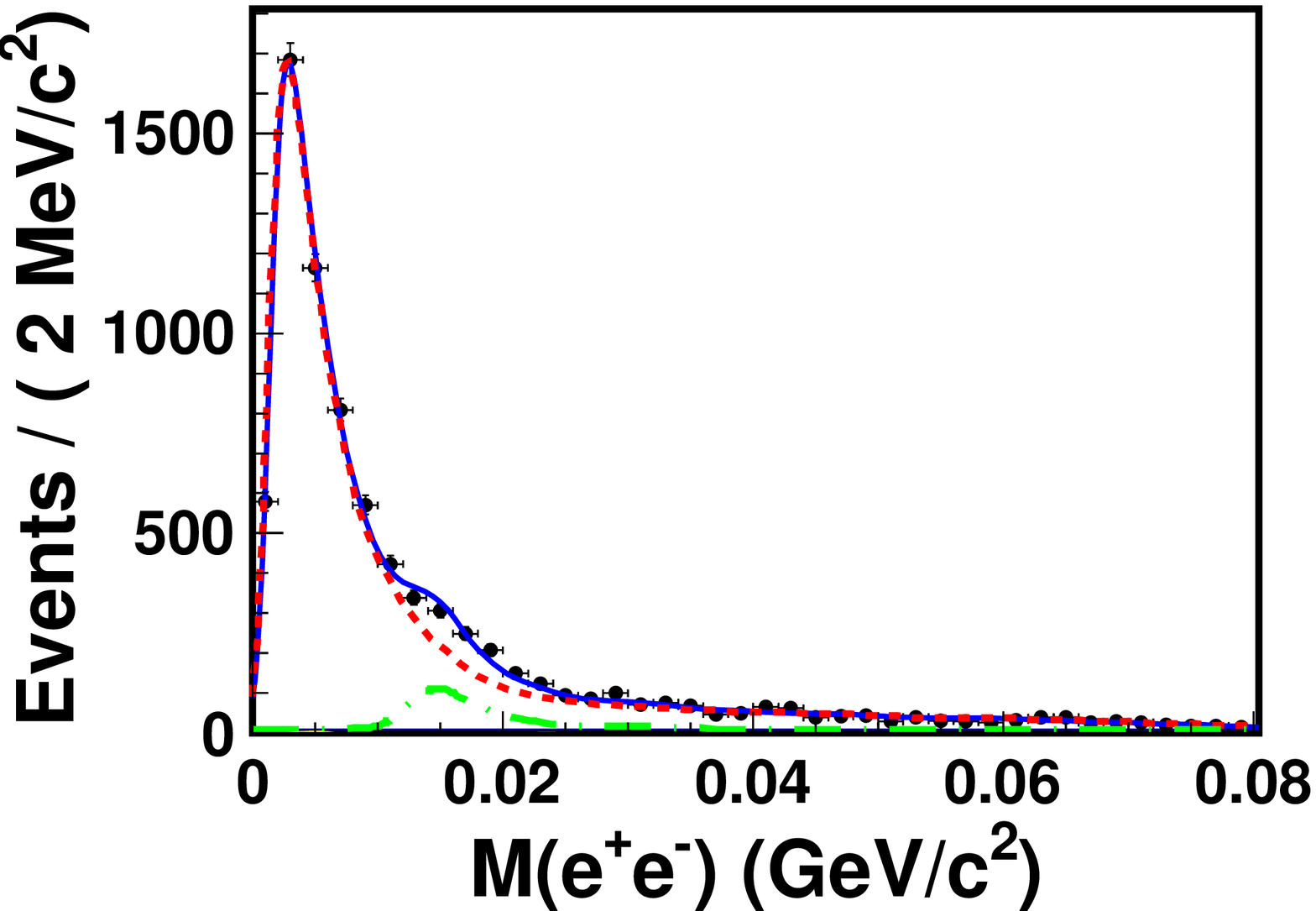}\put(-45,68){\bf \large~(d)}
        \caption{Data of $J/\psi\to\pi^+\pi^-\pi^0, \pi^0\to\gamma e^+e^-$. The distributions of $\pi^0$ masses in (a) and
        (c);  The distributions of the $M_{e^+e^-}$ in (b) and (d).
        The upper two plots [(a) and (b)] are for events without the requirement
        of $\delta_{xy}<2$~cm; the lower two plots [(c) and (d)] are
        for events with the requirement.  The dots with error bars are data. In
        (b) and (d), the (red) dashed curves are the
        MC-simulated signals, the (green) dot-dashed curves are the
        MC-simulated shapes from $J/\psi \to \pi^+\pi^-\pi^0
        (\gamma\gamma)$ in which one of the photons converts to an
        electron-positron pair. Total fits are shown as the (blue) solid
        lines.
        }
    \label{SysConv}
\end{figure}

In this analysis, the peaking background from the
$\gamma$-conversion events in $J/\psi \to P\gamma$ decay is
suppressed by requiring $\delta_{xy}<2$ cm. The uncertainty due to
this requirement is studied using a sample of $J/\psi \to
\pi^+\pi^-\pi^0, \pi^0 \to \gamma e^+e^-$, which includes both the
$\pi^0$ Dalitz decay and $\pi^0 \to \gamma \gamma$ decay with one of
the photons converted to an electron-positron pair.
Figures~\ref{SysConv} (a) and (c) show the $\pi^0$ mass
distributions without and with the requirement, and the purity of
the sample is better than 99\%.  The mass distributions of the
electron-positron pair are shown in Figs.~\ref{SysConv} (b) and (d)
for the events without and with the requirement of $\delta_{xy}<2.0$
cm, respectively. For comparison, the shape of the MC-generated
signal is also plotted. To generate signal events, for the decay  $\pi^0
\to \gamma e^+e^-$, the form-factor is modeled by the simple pole
approximation as:
\begin{eqnarray}\label{simplepole2}
   |F(q^2)|=1+\alpha q^2/m_{\pi^0}^{2},
\end{eqnarray}
where $q$ is the total four-momentum of the electron-positron pair,
$m_{\pi^0}$ is the nominal $\pi^0$ mass, and $\alpha=0.032\pm
0.004$ is the slope parameter~\cite{PDG}.
Extended ML fits to the $M_{e^+e^-}$ distributions are performed
to obtain the signal yields of the $J/\psi \to \pi^+\pi^- \pi^0
(\pi^0 \to \gamma e^+e^-)$ events as shown in Figs.~\ref{SysConv}
(b) and (d).
The data-MC difference of 1.0\% is considered as the systematic
uncertainty for our $\gamma$-conversion veto requiring $\delta_{xy}<2.0$~cm.

The uncertainty from the kinematic fit comes from the inconsistency
between the data and MC simulation of the track helix parameters;
inaccuracies in our MC simulation of photons have previously been shown
to be much smaller ~\cite{GuoYuping}.
Following the procedure described in Refs.~\cite{GuoYuping,
LiaoGuangrui}, we take the difference between the efficiencies with
and without helix parameter corrections as the systematic uncertainty,
which is 1.0\% in each mode.

\begin{figure}[htbp]
    \centering
        \includegraphics[width=7.4cm]{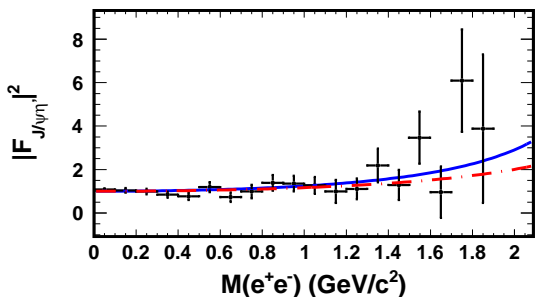}
        \caption{Form factor for $J/\psi \to \etap e^+e^-(\etap \to
        \gamma \pi^+\pi^-)$. The crosses are data, the (red)
        dot-dashed curve is the prediction of the simple pole model
        with the pole mass $\Lambda = m_{\psip} = 3.686$~\GeV, and the fit
        is shown as the (blue) solid curve.
        }
    \label{ttf}
\end{figure}

In the analysis, the form factor is parameterized by the simple
pole approximation as shown in Eq.(\ref{eq:simplepole}) with the
pole mass $\Lambda = m_{\psip}=3.686$~\GeV~ in the signal MC
generator.
Direct information on the pole mass is
obtained by studying the efficiency-corrected signal yields for each
given $M_{e^+e^-}$ bin $i$ for the decay $J/\psi \to \etap
e^+e^-(\etap \to \gamma \pi^+\pi^-)$, which is the channel with the
highest statistics in this analysis. The resolution in $M_{e^+e^-}$
is found to be about 5~MeV in the MC simulation.
This is much smaller than a statistically reasonable bin width,
chosen as 0.1~\GeV, and hence no unfolding is necessary.
The signal yields are background subtracted bin-by-bin and then
efficiency corrected. By using Eq.~(\ref{eq:dgamman}), the value of
the $|F_{J/\psi \etap}|^2$ is extracted for each given bin $i$ as shown
in Fig.~\ref{ttf}.
Fitting this extracted $|F_{J/\psi\etap}|^2$ vs. $M_{e^+e^-}$
data, the pole mass in Eq.(\ref{eq:simplepole})
is determined to be $\Lambda = (3.1\pm1.0)$~\GeV.
To estimate the uncertainty on the signal
efficiency originating from the choice of the pole mass, the signal
events are generated with $\Lambda=3.0$~\GeV~and
$\Lambda=4.0$~\GeV~for each signal mode, respectively.  The relative
difference of the detection efficiency in each signal mode is taken
as the systematic uncertainty, as listed in Table~\ref{sum:sys}.

\begin{table*}[hbtp]
  \caption{Summary of the measurements of the branching fractions, where the first uncertainties are statistical
and the second ones are systematic. The theoretical prediction~\cite{Jinlin} for the branching fractions are listed in the last column.}
  \label{sum:branchratios}\small
  \begin{center}
     \renewcommand{\arraystretch}{1.6}
     \begin{tabular}{cccc}
        \hline\hline
          Mode                     \hspace{1cm}  & \hspace{0.7cm} Branching fraction   \hspace{0.7cm}    &  \hspace{0.7cm}   Combined Result \hspace{0.7cm} &  Theoretical prediction \\ \hline

        $J/\psi\to \etap e^+e^-(\etap \to \gamma\pi^+\pi^-)$           & ~~$(6.01\pm0.20\pm0.34)\times10^{-5}$ & & \\

        $J/\psi\to \etap e^+e^-(\etap\to \pi^+\pi^-\eta)$             & ~~$(5.51\pm0.29\pm0.32)\times10^{-5}$ & $(5.81\pm0.16\pm0.31)\times10^{-5}$  &$(5.66\pm0.16)\times10^{-5}$ \\ \hline

        $J/\psi\to \eta e^+e^-(\eta \to \pi^+\pi^-\pi^0)$      & ~~$(1.12\pm0.13\pm0.06)\times10^{-5}$   &  & \\

        $J/\psi\to \eta e^+e^-(\eta\to \gamma\gamma)$         & ~~$(1.17\pm0.08\pm0.06)\times10^{-5}$  &    $(1.16\pm0.07\pm0.06)\times10^{-5}$ &$(1.21\pm0.04)\times10^{-5}$  \\ \hline

        $J/\psi\to \pi^0 e^+e^-(\pi^0 \to \gamma\gamma)$                     & ~~$(7.56\pm1.32\pm0.50)\times10^{-7}$  &    $(7.56\pm1.32\pm0.50)\times10^{-7}$&$(3.89^{+0.37}_{-0.33}) \times 10^{-7}$ \\
        \hline\hline
      \end{tabular}
  \end{center}
\end{table*}

In the fits to the mass distributions of the pseudoscalar mesons,
the signal shapes are described by the MC signal shape convoluted
with a Gaussian function. Alternative fits are performed by fixing
the signal shape to the MC simulation, and the systematic
uncertainties are set based on the changes observed in the yields.
The uncertainty due to the non-peaking background shape is estimated
by varying the PDF shape and fitting range in the ML fit for each
mode. The changes in yields for these variations give
systematic uncertainties due to these backgrounds.  The numbers of
the expected peaking backgrounds from the photon-conversion in
radiative decay $J/\psi \rightarrow P \gamma$ are summarized in
Table~\ref{gconv}; the errors are negligible for each mode.

The branching fractions for the decay of $\pi^0$, $\eta$ and $\etap$
are taken from the world averages~\cite{PDG}. The corresponding uncertainties
on the branching fractions are taken as the systematic uncertainties. The
uncertainty in the number of $J/\psi$ decays in our data sample is
1.24\%~\cite{Jpsi}, which is taken as a systematic uncertainty.

Assuming all systematic uncertainties in Table~\ref{sum:sys} are independent,
the total systematic uncertainty is obtained by adding them in quadrature.
Totals for the five modes range from 5.4\% to 6.6\%.

\section{Results}

The branching fractions of the EM Dalitz decays $J/\psi\to P
e^+e^-$, where $P$ stands for $\etap$, $\eta$ and $\pi^0$, are
calculated with the following formula:
\begin{eqnarray}
  \label{Br}
  \br{J/\psi \to P e^+ e^-}=
  \frac{N_{S}}{N_{J/\psi}\cdot\br{P\to F}\cdot\epsilon}
\end{eqnarray}
where $N_{S}$ and $\epsilon$ are the number of signal events and the
detection efficiency for each mode, respectively, listed in Table~\ref{yields}.
Here, $N_{J/\psi} = (225.3\pm2.8)\times 10^{6}$ is the number of
$J/\psi$ events, and $\br{P\to F}$ is the product of the branching fraction of the
pseudoscalar decays into the final states $F$, taken from the
PDG~\cite{PDG}. The calculated branching fractions are summarized in
Table~\ref{sum:branchratios}.

The branching fractions of $J/\psi \to \etap e^+e^-$ and $J/\psi \to
\eta e^+e^-$ measured in different decay modes are consistent with
each other within the statistical and uncorrelated systematic uncertainties.
In Table~\ref{sum:sys}, the items with asterisks denote the correlated systematic
errors while the others uncorrelated. The measurements from different modes
are therefore combined with the approach in Ref.~\cite{combined}, which uses
a standard weighted least-squares procedure taking into consideration the
correlations between the measurements. For $J/\psi\to \etap e^+e^-$, the
correlation coefficient between $\etap\to\gamma\pi^+\pi^-$ and $\etap\to\pi^+\pi^-\eta$ is
$\rho(1,2) = 0.46$; for $J/\psi\to\eta e^+e^-$, it is
$\rho(1,2)=0.13$. The weighted averages of the BESIII measurements
are listed in Table~\ref{sum:branchratios}.

\section{Summary}

In summary, with a sample of $(225.3\pm2.8)\times 10^{6}$ $J/\psi$
events in the BESIII detector,  the EM Dalitz decays $J/\psi \to P
e^+e^-$, where $P$ stands for $\etap$, $\eta$ and $\pi^0$,  have
been observed for the first time. The branching fractions of
$J/\psi\to \etap e^+e^-$, $J/\psi \to \eta e^+e^-$ and $J/\psi \to
\pi^0 e^+e^-$ are measured to be: $\mathcal{B}(J/\psi\to \etap
e^+e^-)=(5.81\pm0.16\pm0.31)\times10^{-5}$,  $\mathcal{B}(J/\psi\to \eta
e^+e^-)=(1.16\pm0.07\pm0.06)\times10^{-5}$ and $\mathcal{B}(J/\psi\to \pi^0
e^+e^-)=(7.56\pm1.32\pm0.50)\times10^{-7}$, respectively. The measurements
for  $J/\psi\to \etap e^+e^-$ and $J/\psi \to \eta e^+e^-$ decay
modes are consistent with the theoretical prediction in
Ref.~\cite{Jinlin}. However, the theoretical prediction for the decay
rate of $J/\psi \to \pi^0 e^+e^-$ based on the VMD model is
$(3.89^{+0.37}_{-0.33}) \times 10^{-7}$, about 2.5 standard
deviations from the measurement in this analysis, which may indicate
that further improvements of the QCD radiative and relativistic
corrections are needed.

\section{ACKNOWLEDGMENT}

The BESIII collaboration thanks the staff of BEPCII and the
computing center for their strong support. The authors thank Mao-Zhi
Yang for useful discussions. This work is supported in part by the
Ministry of Science and Technology of China under Contract No.
2009CB825200; Joint Funds of the National Natural Science Foundation
of China under Contracts Nos. 11079008, 11179007, 11179014, U1332201;
National Natural Science Foundation of China (NSFC) under Contracts
Nos. 10625524, 10821063, 10825524, 10835001, 10935007, 11125525,
11235011; the Chinese Academy of Sciences (CAS) Large-Scale
Scientific Facility Program; CAS under Contracts Nos. KJCX2-YW-N29,
KJCX2-YW-N45; 100 Talents Program of CAS; German Research Foundation
DFG under Contract No. Collaborative Research Center CRC-1044;
Istituto Nazionale di Fisica Nucleare, Italy; Ministry of
Development of Turkey under Contract No. DPT2006K-120470; U. S.
Department of Energy under Contracts Nos. DE-FG02-04ER41291,
DE-FG02-05ER41374, DE-FG02-94ER40823, DESC0010118; U.S. National
Science Foundation; University of Groningen (RuG) and the
Helmholtzzentrum fuer Schwerionenforschung GmbH (GSI), Darmstadt;
WCU Program of National Research Foundation of Korea under Contract
No. R32-2008-000-10155-0.

\end{document}